\documentclass[aip,twocolumn,jcp]{revtex4}
\usepackage{graphicx}
\usepackage{amssymb}
\usepackage{amsmath}
\usepackage{braket}
\usepackage{color}

\newcommand{\R}{\mathbf{r}}

\begin{document}

\title{Subsystem density functional theory with meta generalized gradient approximation exchange-correlation functionals}

\author{Szymon \'Smiga$^a$, Eduardo Fabiano$^{b,c}$, Savio Laricchia$^d$, Lucian A. Constantin$^c$ and Fabio Della Sala$^{b,c}$}
\affiliation{$^a$ Institute of Physics, Faculty of Physics, Astronomy and Informatics, Nicolaus Copernicus University, Grudziadzka 5, 87-100 Torun, Poland}
\affiliation{$^b$Istituto Nanoscienze-CNR, Euromediterranean Center for Nanomaterial Modelling and Technology (ECMT), Via per Arnesano 16, I-73100 Lecce, Italy}
\affiliation{$^c$Center for Biomolecular Nanotechnologies@UNILE, 
Istituto Italiano di Tecnologia (IIT), Via Barsanti, 73010 Arnesano (LE), Italy}
\affiliation{$^d$Department of Physics, Temple University, 
Philadelphia, Pennsylvania 19122, USA}

\begin{abstract}
We analyze the methodology and the performance of subsystem density functional
theory (DFT) with meta-generalized gradient approximation (meta-GGA) 
exchange-correlation functionals for non-bonded systems.
Meta-GGA functionals depend on the Kohn-Sham kinetic energy density (KED), 
which is not known as an explicit functional of the density.
Therefore, they cannot be directly applied in subsystem DFT calculations. 
We propose a Laplacian-level approximation to the KED which overcomes the
problem and provides a simple and accurate way to apply meta-GGA exchange-correlation 
functionals in subsystem DFT calculations. 
The so obtained density and energy errors, with respect to the 
corresponding supermolecular calculations, are comparable
with conventional approaches, depending almost exclusively on the 
approximations in the non-additive kinetic embedding term.
An embedding energy error decomposition explains the accuracy of our method.
\end{abstract}

\maketitle

\section{Introduction}
Subsystem density functional theory \cite{cortona,wesoiter,wesorev,neugrev10,subdftrev,pavanello15} is nowadays
attracting increasing interest in the density functional
theory (DFT) \cite{hk,ks} community. This is due to its
promise of achieving potentially exact results at 
a reduced computational cost,
as well as to
the high insight into the nature of interacting
systems provided by the associated embedding potentials.
{Thus, numerous applications related to non-covalent 
\cite{wesolowski96fhnch,weso97,Wesolowski98,wesotran03,kevorkyants06hyd,dulak07,dulak07geom,lastra08,gotz09,fradelos11,apbe,apbek,fde_lhf,fde_hyb_ene,fdefract,fde_ct,fde_lap,schluns05,pavanellovdw}
as well as covalent bonded systems \cite{jacob08prot,fux08,beyhan10,fux10} have been considered.
In addition, the frozen density embedding (FDE)
method \cite{wesorev,wesowarh93,hodak08} has emerged as a practical tool for
efficient simulations of different properties \cite{neugebauer05solv,jacob06solv,neugebauer05spin,hodak08,kaminski10,fradelos11,kiewisch13}.
We also recall that the FDE method with the iterative procedure of Ref. \onlinecite{wesoiter} is a computational implementation of the subsystem DFT.
}

However, the accuracy of subsystem DFT calculations is
practically limited by two factors.
First, the term describing the interaction energy between different
subsystems depends on the non-additive kinetic energy (KE),
which must be described by an explicit density functional.
Second, the embedding potential, which is required to
describe the mutual interaction between the subsystems,
must be a local multiplicative potential. Thus, it can
contain only local or semilocal approximations for the
non-additive contribution of the kinetic and 
exchange-correlation (XC) terms to the embedding potential.
%
{Nevertheless, both limiting factors have currently
been, at least partially, overcome. In fact, past years have seen 
the development of numerous
KE functionals which can be suitable for
subsystem DFT: GGA functionals \cite{apbe,apbek,tran02,llp91,lc94,thak92}, Laplacian-level meta-GGA functionals \cite{fde_lap} and non-decomposable
approach \cite{lastra08}.
For a recent review of all KE functionals see Ref. \onlinecite{weso_chap}.
}
{Moreover, several works have extended}
the subsystem formulation of DFT beyond
the conventional Kohn-Sham (KS) framework, considering e.g., 
hybrid functionals \cite{fde_hybrid},
{embedded interacting wave functions \cite{weso08det},}
orbital-dependent effective exact exchange methods
\cite{fde_lhf,jacob05dip}, or density matrix \cite{pernalweso09}.
Nevertheless, to date, no attempt has been made to 
consider subsystem DFT calculations using
meta generalized gradient approximation (meta-GGA)
XC functionals.

A meta-GGA XC functional is defined by the general formula
\begin{equation}\label{e1}
E_{xc} = \int e_{xc}(\rho(\R),\nabla\rho(\R),\tau^\text{KS}(\R))d\R\ ,
\end{equation}
where $e_{xc}$ is the  XC energy density,
$\rho$ is the electron density and 
\begin{equation}\label{tau_eq}
\tau^\text{KS}(\R) = \frac{1}{2}\sum_i^\mathrm{occ.}\left|\nabla\phi_i^\text{KS}(\R)\right|^2
\end{equation}
is the positive-defined KS kinetic-energy density (KED), with
$\phi_i^\text{KS}$ being the occupied KS orbitals of the system.
Meta-GGAs are attracting increasing popularity
\cite{tpss,revtpss,js,tpssloc,bloc,vsxc,477481,m06l,m11l,regtpss,mggms_1,mgga_ms2,sunpnas,beefmeta14,b97mv2015}
because they can satisfy numerous exact constraints
of the XC energy \cite{tpss,revtpss,tpssloc,bloc_hole,fkato}, 
achieve a remarkable level of accuracy
\cite{bloc,vsxc,477481,m06l,m11l,xiao13,sunprl,stare04,adam00,riley07,hammer12,sunpnas,m06surf,sunmeta11,hao13,dihyd14}, and
describe excitonic effects in crystals \cite{vignale}.
In short, meta-GGA functionals has much larger accuracy/computational cost than GGAs, and thus should be 
preferred to the latter. 

However, $\tau^\text{KS}$ is not an explicit functional of the 
density, thus the implementation of meta-GGA functionals
within the conventional KS scheme is not straightforward
\cite{arbuznikov03}, since it requires the
calculation of $\delta\tau^\text{KS}/\delta\rho$.
Thus, meta-GGAs are often implemented within
a Generalized Kohn-Sham scheme (GKS) \cite{seidl96}.
For this reason, so far, meta-GGA XC functionals
have never been employed in
subsystem DFT calculations.

In this paper we consider this issue and develop
the theory and the methodology required to
perform subsystem DFT calculations at the meta-GGA level.
In particular, we consider proper semilocal
approximations for the XC embedding contributions 
and test them on a  set of non-covalent
complexes assessing the accuracy of
the resulting energies and densities.

Thus, the paper is organized as follows: in section \ref{sec:theor} we present the general theory
for subsystem DFT with arbitrary orbital-dependent XC functionals 
(thus including both meta-GGA and hybrid functionals) as well as 
different schemes to approximate the KED;
Computational details are reported in section \ref{sec:compdet};
Results for non-covalent complexes are presented in section \ref{sec:res}.
Finally, in Section \ref{concl} we summarize our conclusions.

\section{Theory}\label{sec:theor}
\subsection{Subsystem DFT with arbitrary orbital-dependent XC functionals}
 \label{sec:ta}
Within the subsystem formulation of density
functional theory a given system is partitioned into 
two subsystems $A$ and $B$, each defined by its
nuclear potential $v_\text{nuc}^A(\R)$ and $v_\text{nuc}^B(\R)$,
respectively. Accordingly, 
{the electron density of the total system is constructed} 
as $\rho_{A+B}=\rho_A+\rho_B$,
where the two subsystem densities integrate to
$N_A$ and $N_B$, respectively.
 For simplicity we focus here on the case where $N_A$ and $N_B$ are
integer numbers.

The ground-state solution of the problem is
given by the set of coupled equations
\begin{eqnarray}
\label{e3}
\frac{\delta F_\text{HK}[\rho_A]}{\delta\rho_A(\R)} + v^A_\text{emb}(\R) + v^A_\text{nuc}(\R)  & = & \mu \\
\label{e4}
\frac{\delta F_\text{HK}[\rho_B]}{\delta\rho_B(\R)} + v^B_\text{emb}(\R) + v^B_\text{nuc}(\R)  & = & \mu\ ,
\end{eqnarray}
where $F_\text{HK}$ is the universal functional of
Hohenberg and Kohn, $\mu$ is the chemical potential
{(which is, in the case of the exact 
theory, equal in the two subsystems and equal to the supramolecular one \cite{fdefract,pavanello15,gritchap})}
and
\begin{equation}\label{e5}
v_\text{emb}^I(\R) = \frac{\delta F_\text{HK}[\rho_I+\rho_{II}]}{\delta\rho_I(\R)} - \frac{\delta F_\text{HK}[\rho_I]}{\delta\rho_I(\R)} + v_\text{ext}^{II}(\R)
\end{equation}
is the embedding potential \cite{fde_hybrid}
with $I=A,B$ and $II=B,A$, respectively
(this convention will be used throughout).

In this work we consider for $F_\text{HK}$ the following partition:
\begin{equation}
F_\text{HK}[\rho] = F_s[\rho] + J[\rho]
\end{equation}
with
\begin{equation}
      F_s[\rho]  = 
\min_{\Phi\rightarrow\rho}\left\{\langle\Phi|\hat{T}|\Phi\rangle+ E_{xc}[\{\phi_i\}]\right\} 
\label{eq:gen}
\end{equation}
where $\rho$ is any $N$-representable electron density,
$J$ is the classical Coulomb energy, 
$\hat{T}$ is the KE operator,
$E_{xc}$ is a proper orbital-dependent XC functional (e.g. in the form given by Eq. (\ref{e1})),
$\Phi$ is a Slater determinant, and $\{\phi_i\}$ 
are its single-particle orbitals.
Equation (\ref{eq:gen}) is quite general and includes not only meta-GGA functionals, but all the orbital-dependent ones.
Then, Eqs. (\ref{e3}) and (\ref{e4}) become
\begin{equation}\label{e7}
\frac{\delta F_{s}[\rho_I]}{\delta\rho_I(\R)} + \frac{\delta J[\rho_I]}{\delta\rho_I(\R)} + v^I_\text{emb}(\R) + v^I_\text{nuc}(\R) = \mu \;\; ,
\end{equation}
and the embedding potential of Eq. (\ref{e5}) can be written as
\begin{equation}\label{embpot}
v_\text{emb}^I(\R) = 
v_\text{ext}^{II}(\R) +  \frac{\delta J[\rho_{II}]}{\delta\rho_{II}(\R)} +\frac{\delta F_{s}^\text{nadd}[\rho_I,\rho_{II}]}{\delta\rho_I(\R)} 
\end{equation}
where 
\begin{equation}\label{e18}
F_s^\text{nadd}[\rho_A,\rho_B] = F_s[\rho_A+\rho_B] - F_s[\rho_A] - F_s[\rho_B]
\end{equation}
is the non-additive contribution to the kinetic plus XC energy and 
we used the fact that the Coulomb potential is additive. 

At this point, following the GKS scheme
\cite{seidl96,fde_hybrid},
we introduce, for each subsystem (for example $I$)
an auxiliary system of particles having 
the following properties: 
(i)
it has the same 
ground-state density as our original embedded subsystem $I$;
(ii)
 it is described by a single Slater determinant $\Phi^I$;
(iii) 
the ground-state energy is the minimum of the energy functional
\begin{equation}
E[\Phi^I] = \langle\Phi^I|\hat{T}|\Phi^I\rangle + 
E_{xc}[\{\phi^I_i\}] + \int w^I(\R)\rho_I(\R)d^3\R\ ,
\end{equation}
where $w$ is a (yet unknown) external local potential.
 
The ground-state energy of this system is defined
via the constrained search procedure as
\begin{eqnarray}
E_0 &=& \min_{\rho_I\rightarrow N_I}\left\{\min_{\Phi^I\rightarrow\rho_I}
E[\Phi^I]\right\} \nonumber \\
&=& \min_{\rho_I\rightarrow N_I}\left\{
F_s[\rho_I] + \int w^I(\R)\rho_I(\R)d^3\R
\right\} \, .
\end{eqnarray}
Hence, the ground-state is described by the Euler equation
\begin{equation}\label{e10}
\frac{\delta F_{s}[\rho_I]}{\delta\rho_I(\R)} + w^I(\R) = \mu\ .
\end{equation}
Comparing Eqs. (\ref{e7}) and  (\ref{e10})
and making use of the property (i) we thus find
\begin{equation}
\label{eqww}
w^I(\R) = v^I_\text{nuc}(\R)+ \frac{\delta J[\rho_I]}{\delta\rho_I(\R)} + v^I_\text{emb}(\R) \, .
\end{equation}

On the other hand, properties (ii) and (iii) imply
directly that
the ground-state of the auxiliary system is described by the set of 
single-particle equations
\begin{equation}
\label{eqddd}
\left[-\frac{1}{2}\nabla^2 + w^I(\R)\right]\phi_i^I(\R) 
+ \frac{1}{2}\frac{\delta E_{xc}[\{\phi^I_i\}]}{\delta\phi^I_i(\R)} = \epsilon^I_i\phi^I_i(\R) \ ,
\end{equation}
where we assumed real orbitals and $\epsilon^I_i$ are Lagrange multipliers to ensure orbital orthonormality.
For a meta-GGA functional the
last term on the left hand side can be evaluated as \cite{Arbuznikov2003495,B207171A}
\begin{equation}
\frac{\delta E_{xc}}{\delta\phi_i} = 2\left[ \frac{\partial e_{xc}}{\partial \rho}\phi_i 
+ \left(\nabla\cdot\frac{\partial e_{xc}}{\partial \nabla\rho}\right)\phi_i 
- \frac{1}{2} \nabla\cdot\left(\frac{\partial e_{xc}}{\partial \tau}\nabla\phi_i\right)\right]\ .
\end{equation}
Combining Eq. (\ref{eqww}) with Eq. (\ref{eqddd}) the operational equations to solve the subsystem
ground-state problem are, finally:
\begin{eqnarray}
&&\left[-\frac{1}{2}\nabla^2 + \frac{\delta J[\rho_A]}{\delta\rho_A(\R)} 
+ v^A_\text{emb}(\R) 
+ v^A_\text{nuc}(\R)\right]\phi^A_i(\R) + \nonumber \\
&& \quad\quad\quad\quad 
+ \frac{1}{2} \frac{\delta E_{xc}[\{\phi^A_i\}]}{\delta\phi^A_i(\R)} 
= \epsilon^A_i\phi^A_i(\R) \label{eqscfa} \\
&&\left[-\frac{1}{2}\nabla^2 + \frac{\delta J[\rho_B]}{\delta\rho_B(\R)} 
+ v^B_\text{emb}(\R) 
+ v^B_\text{nuc}(\R)\right]\phi^B_i(\R) + \nonumber \\
&& \quad\quad\quad\quad 
+ \frac{1}{2}\frac{\delta E_{xc}[\{\phi^B_i\}]}{\delta\phi^B_i(\R)} 
= \epsilon^B_i\phi^B_i(\R) \label{eqscfb}
\end{eqnarray}
with the embedding potential $v^I_\text{emb}$ given by Eq. (\ref{embpot}). 
{ Note that, if the embedding potential is treated exactly, Eqs. (\ref{eqscfa}) and (\ref{eqscfb}) admit in general multiple solutions
\cite{humbert13,pavanello15,wassermanrev}.}
Once the orbitals have been obtained, the total
electron density is computed as
\begin{equation}
\rho_{A+B} (\R)= \sum_{i=1}^{N_A}\left|\phi_i^A(\R)\right|^2 
+ \sum_{i=1}^{N_B}\left|\phi_i^B(\R)\right|^2\ ,
\end{equation}
whereas the total electronic energy is calculated as
\begin{eqnarray}
\nonumber
 &&E_{A+B}[\rho_A,\rho_B] = 
\langle\Phi^A|\hat{T}_s|\Phi^A\rangle+ 
\langle\Phi^B|\hat{T}_s|\Phi^B\rangle  \\
\nonumber
&& + E_{xc}[\{\phi_i^A\}] + E_{xc}[\{\phi_i^B\}] + J[\rho_A+\rho_B] + \\
\nonumber
&& + \int \left(\rho_A(\R)+\rho_B(\R)\right)\left(v_\text{nuc}^A(\R)+v_\text{nuc}^B(\R)\right)d^3\R +F_s^\text{nadd}[\rho_A,\rho_B]\ . \label{eq:ener}
\end{eqnarray}
%
%
%

Note that the formalism introduced above is quite general and can be applied to 
GGA, hybrid-GGA, and meta-GGA functionals.

\subsection{Non-additive embedding contributions}
To perform embedding calculations according to
the theory detailed in Sec. \ref{sec:ta}
we need to compute the non-additive contribution $F_s^\text{nadd}[\rho_A,\rho_B]$
(see Eq. (\ref{eq:ener})) 
and its derivatives with respect to $\rho_A$ and $\rho_B$ (see Eq. (\ref{embpot})).
This is not an easy task for two main reasons \cite{fde_hybrid}: 

(i)
The computation of $F_s[\rho_A+\rho_B]$ requires the knowledge of the ground-state Slater determinant of the 
total system, $\Phi^{A+B}$, which is not available by definition in a subsystem
calculation. It can only be obtained by an inverse KS
\cite{zhao94,wuyang03,weso12} 
calculation starting from the ground-state total density $\rho_{A+B}$ 
(for some examples within FDE see Refs. 
\onlinecite{roncero08,roncero09,fux10,goodpaster10,goodpaster11}).

(ii)
The $F_s$ functional depends explicitly on the orbitals and
has only an implicit dependence on the density. Therefore,
to make the required functional derivatives special techniques
are required, such as the optimized effective potential method \cite{kummel2008,jacob11,hesselman,staroverov,heatonburgess}.

Consequently, in search of a practical computational procedure to perform
subsystem DFT calculations with meta-GGA XC functionals, we 
propose to consider, in analogy with Refs. \onlinecite{fde_hybrid,fde_lhf},
in Eq. (\ref{e18}) the semilocal approximation
\begin{eqnarray}
F_s[\rho]       &\approx& \widetilde{F}_s[\rho]       = \widetilde{T}_s[\rho]       + \widetilde{E}_{xc}[\rho]\ ,  \label{eq:appfs} \\
F_s^{nadd}[\rho] &\approx& \widetilde{F}_s^{nadd}[\rho] = \widetilde{T}_s^{nadd}[\rho] + \widetilde{E}_{xc}^{nadd}[\rho]\ , \label{eq:appfsnadd}
\end{eqnarray}
where the tilde  denotes that an approximated (semilocal)
functional of the density is used.

For the kinetic term standard semilocal approximations can
be employed 
\cite{apbe,apbek,tran02,lc94,thak92,weso_chap}.
For the XC part, instead, two main possibilities can be envisaged:
\begin{itemize}
\item[i)] As a first simple option it is possible to use for
$\widetilde{E}_{xc}$ the GGA functional ``most similar'' 
to the meta-GGA functional used for subsystems calculations.
For example, if the TPSS \cite{tpss} functional is used 
for subsystems calculations, 
the natural choice will be to use the PBE functional
for the non-additive contribution. 
In fact, the TPSS functional has been constructed 
as an extension of the PBE functional.
This choice resembles that of Refs. 
\onlinecite{fde_hybrid,fde_lhf} and may be
expected to yield reasonable results.
The accuracy of this combination will be verified in section \ref{sec:res}.

\item[ii)] A second, possibly better, choice is to
retain for $\widetilde{E}_{xc}$ the meta-GGA
form, but replacing $\tau$ with a suitable semilocal approximation.
We recall that this is in general a very hard task \cite{yang86,alva07}.
However, for our purposes we only need the non-additive XC energy and,
as shown in section \ref{sec:res},
a large error cancellation effect can thus be expected.
\end{itemize}

\subsection{Semilocal models for the kinetic energy density} \label{sec:mods}
%
{In this subsection we consider the construction of
semilocal models for the KED. However, since the
KED is not an observable, it is defined only up
to a gauge integrating to zero (and vanishing in the functional derivative).
Thus, to fix our working definition of the KED
we decide to consider here only the positive-defined KED
(Eq. (\ref{tau_eq}); see Ref. \onlinecite{ayers02} for a discussion on this topic), which is also the most commonly used in meta-GGA XC
functionals.}

To model the positive-defined KED in
our subsystem DFT calculations we consider the following two
semilocal approximations:
\begin{eqnarray}
\label{tau1}
\tilde{\tau} & \approx & \tau^\text{TF} + \tau^{W} \doteq \tau^1 \\
\label{tau2}
\tilde{\tau} & \approx &  \tau^\text{revAPBEK} + \frac{20}{9}\tau^\text{TF}q \doteq \tau^L,
\end{eqnarray}
where $\tau^\text{TF}=(3/10)(3\pi^2)^{2/3}\rho^{5/3}$ is the
Thomas-Fermi (TF) kinetic energy density \cite{thomas26,fermi28,fermi27},
$\tau^W=(5/3)\tau^{\text{TF}}s^2 = |\nabla\rho|^2/(8\rho)$ 
is the von Weizs\"acker kinetic energy density \cite{vw}
(with $s=|\nabla\rho|/[2(3\pi^2)^{1/3}\rho^{4/3}]$ the reduced gradient),
$\tau^\text{revAPBEK}$ is the revAPBEK kinetic energy density
\cite{apbe,apbek}, and $q=\nabla^2\rho/[4(3\pi^2)^{2/3}\rho^{5/3}]$
is the reduced Laplacian.

The $\tau^1$ model of Eq. (\ref{tau1}) is a simple GGA model that is exact for  
a uniform density perturbed by a small-amplitude short-wavelength density wave and motivated by
the basic requirements that  $\tau\approx\tau^\text{TF}$ in the slowly-varying
density limit and $\tau\approx\tau^W$ \cite{alpha}
in tail regions (where $\tau^{\text{TF}}\rightarrow 0$)  and iso-orbital regions.
Moreover, this simple model fulfills the important constraint
$\tau^W\leq\tau$, i.e. that $0 \leq z\leq 1$, with $z=\tau^{W}/\tau$, 
which has actually been used in the construction of several 
meta-GGA XC functionals \cite{tpss}.

The $\tau^L$ model of Eq. (\ref{tau2}) is a Laplacian-level
meta-GGA model and requires several considerations:
\begin{itemize}
\item[i)]  We first recall that
$(20/9)\tau^\text{TF}q=(1/6)\nabla^2\rho$.
Thus, the kinetic energy and potential corresponding to $\tau^L$ are
identical with the revAPBEK ones 
(recall that a term proportional to the Laplacian of the density does not 
contribute to the energy and the potential).
If the revAPBEK functional is used 
for $\tilde{T}_s$ and $\tilde{T}_s^{nadd}$ 
in Eqs. (\ref{eq:appfs}) and (\ref{eq:appfsnadd}), then
the revAPBEK KE approximation is ``de facto'' 
the only functional approximation used in the subsystem DFT 
meta-GGA calculation.
{In fact, in our implementation we 
 use the same KE functional in both  $\tau^L$ and  $\widetilde{T}_s^{nadd}[\rho]$.}
\item[ii)] The term containing $q$ is  of fundamental importance to reproduce 
the correct KE density \cite{yang86,alva07}).
In  Eq. (\ref{tau2}) the coefficient of the reduced 
Laplacian term comes from the lowest-order Laplacian contribution
to the second-order gradient expansion of the KE \cite{brack1976,kirzhnits57}.
Other coefficients could be used as well \cite{yang86,alva07}, 
but we found that the non-empirical one in  Eq. (\ref{tau2}) 
is quite accurate for our purpose, 
even if  Eq. (\ref{tau2})  is not exact in the asymptotic region 
(see Appendix \ref{appa}).
\item[iii)] 
The revAPBEK functional, which is used in the definition of $\tau^L$, 
recovers by construction the
modified second-order gradient expansion of the KE\cite{mge2}, which was constructed from semiclassical theory of atoms.
Actually, the revAPBEK functional, 
{which does not contain any 
empirical parameter fitted on kinetic energies}, has been found to be very accurate 
for the description of non-covalent complexes within subsystem DFT 
\cite{apbek}. 
{However, the accuracy of the $\tau^L$ model is not
mandatorily related to the use of revAPBEK: other state-of-the-art GGA \cite{apbe,apbek,tran02,llp91,lc94,thak92} and
meta-GGA \cite{fde_lap} KE approximations can be expected to yield similar accuracy.}

\item[iv)] The Laplacian term will diverge at the core ($q \rightarrow -\infty$). 
However, such a bad behavior in the core region is not a problem for subsystem DFT
calculations, since for the non-additive XC energy and potential 
these contributions cancel almost completely.
This cancellation is shown in Figure 6 of 
Ref. \onlinecite{fde_lap} where a reduced gradient and Laplacian decomposition
of the non-additive KE is reported, showing that
 only small values of $q$ contribute significantly.
\end{itemize}

To show the physical significance of the two models introduced above and
offer a preliminary test of the expectable performance,
we consider their application to the test case of the Ne dimer,
an example for weakly interacting systems.
As meta-GGA exchange correlation functional we consider 
the TPSS \cite{tpss} one.
This is in fact one of the first and most popular non-empirical 
meta-GGAs and is used here as an exemplary case.

\begin{figure}
\includegraphics[width=\columnwidth]{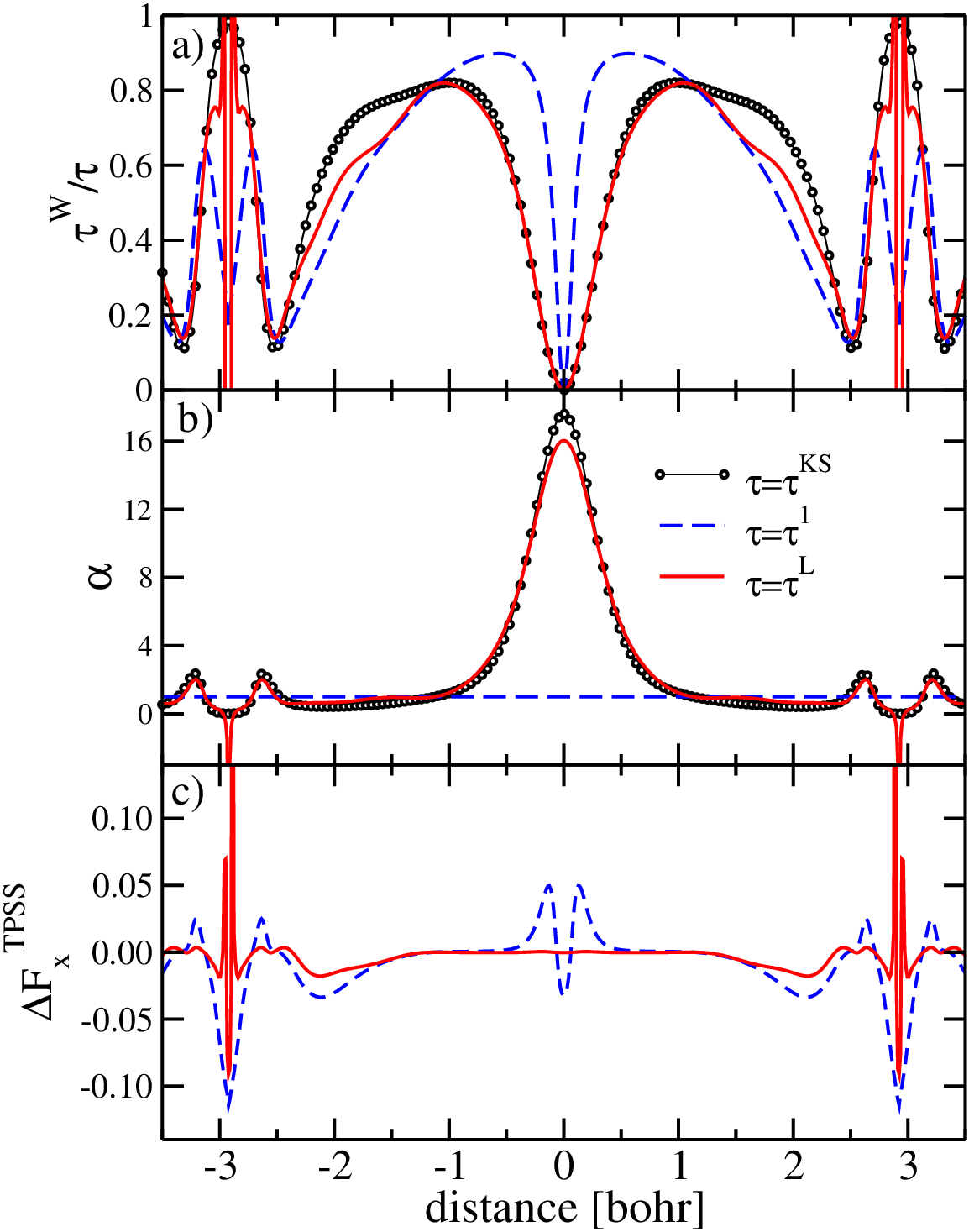}
\caption{\label{nenediff} 
Plot of $z=\tau^\text{W}/\tau$ (panel a), $\alpha=(\tau-\tau^W)/\tau^{\text{TF}}$ (panel b) 
and $\Delta F_x^\text{TPSS}$ (panel c) for the Ne dimer along the main axis 
computed using the kinetic energy density models of Eqs. (\ref{tau1}) and (\ref{tau2}).}
\end{figure}

In Fig. \ref{nenediff}a we report the 
quantity $z=\tau^W/\tau$ for the two KE density approximations 
Eqs. (\ref{tau1}) and (\ref{tau2})
as a function of the distance $d$ along the dimer axis.
In the middle of the bond ($d=0$)
all curves go to zero, because the gradient of the density 
and thus $\tau^\text{W}$ vanish.
The $\tau^1$ model is accurate only at few points: $|d|\approx 1$ bohr 
and near the core. However, 
exactly at the nucleus position the exact $\tau^W/\tau$ is close to 1 
\cite{alpha}; on the other hand at this point
the density is very large so that $\tau^\text{TF} \gg \tau^\text{W}$, 
thus $\tau^W/\tau^1$ is much less than 1.
The Laplacian model $\tau^L$ is everywhere (but in the core) much more accurate than  $\tau^1$. In particular, it reproduces almost exactly
$\tau^\text{KS}$  in the region $|d| \le 1.5$ bohr.

Alternatively, in Fig.  \ref{nenediff}b we report the quantity
\begin{equation}
\alpha=\frac{\tau-\tau^W}{\tau^{\text{TF}}}=\frac{5}{3}s^2 \left(\frac{1}{z}-1\right)
\end{equation}
which is the main, direct ingredient of the meta-GGA (TPSS) exchange enhancement factor  and thus on the calculation of
the exchange energy.
Differences about the various approaches are now more evident. For the $\tau^1$ model $\alpha=1$ everywhere in the space (by construction): 
this can be a good approximation only for slowly varying densities.
Instead, for molecular systems, the exact alpha shows significant oscillations and it is very high  ($\alpha\approx17.6$) in the bond. 
These oscillations are correctly reproduced by the  $\tau^L$ model, which yields a very accurate value at the bond ($\alpha\approx 16.0$).
 $\tau^L$ is significantly different from the exact one only near the core, where it is actually negative. 
Finally,  In Fig.  \ref{nenediff}c we report the quantity
\begin{equation}
\Delta F_x^\text{TPSS}= F_x^\text{TPSS}(\rho,\nabla\rho,\tilde{\tau})- F_x^\text{TPSS}(\rho,\nabla\rho,\tau^{KS})
\end{equation}
where $F_x^{TPSS}$ is the TPSS exchange enhancement factor \cite{tpss}, so that
\begin{equation}
E_x^\text{TPSS}=\int \rho \epsilon_{x}^\text{LDA}(\rho) F_x^\text{TPSS}(\rho,\nabla\rho,\tau^{KS}) d\R\ ,
\end{equation}
where $\epsilon_{x}^\text{LDA}$ is the LDA exchange energy per particle.
The quantity $\Delta F_x^\text{TPSS}$ indicates how much 
the approximation in $\tilde{\tau}$ will impact on the accuracy of the exchange energy.

The plots in  Fig. \ref{nenediff}b confirm the high accuracy of 
the $\tau^L$  model, whereas the  $\tau^1$ model shows 
quite larger differences  in the region $|d|< 0.5$ au.

In section \ref{sec:res} we will consider the  performance of the two models 
for subsystem DFT calculations.

\section{Computational details}\label{sec:compdet}
To assess the possibility of performing subsystem DFT calculations
using meta-GGA functionals we carried on test simulations
on different non-covalent complexes.
To this end, for simplicity, we considered as meta-GGA XC functional 
the TPSS one \cite{tpss}. 
Other meta-GGA XC functionals will be considered in detail in a future publication.

In our calculations we used 
different approximations for the non-additive XC terms.
We refer to each of these using the notation method1/method2,
that denotes that method1 was used for the subsystems and
method2 to compute the non-additive XC contribution.
In more details:
\begin{enumerate}
\item As a simple choice we computed the non-additive XC contributions using the 
PBE XC functional \cite{pbe}, i.e. we set $\widetilde{E}_{xc}=E_{xc}^{PBE}$.
This approach is labeled TPSS/PBE.
\item Alternatively, we computed the non-additive XC terms using the XC TPSS
functional but using the $\tau^1$ model of Eq. (\ref{tau1})
to mimic the positive-defined KED.
This approach is named TPSS/TPSS-1.
\item Finally, we considered the same case as above but using the  $\tau^L$ model
of Eq. (\ref{tau2}). This approach is labeled TPSS/TPSS-L.
Note that in this case, because of the negative divergence of $q$, 
the model we use for $\tau$ is not 
guaranteed to respect the bound $\tau^W\leq\tau$. 
Nevertheless, the TPSS functional is numerically well 
defined also for $z<0$ or $z>1$; moreover these values 
occur only near the core which is not relevant for non-additive contributions, 
see point iv) of section \ref{sec:mods}.
\end{enumerate}
We remark that the first two approximations are GGA ones, while the last
one is a Laplacian-level meta-GGA method.
For comparison also subsystem DFT calculations using the PBE
\cite{pbe} and PBE0 \cite{pbe0_1,pbe0_2} XC functionals 
were considered. The former one requires no approximations
for the non-additive XC terms and implements the PBE/PBE approach; 
the latter one instead uses a semilocal approximation as described 
in Ref. \onlinecite{fde_hybrid} and yields the PBE0/PBE approach.

In all calculations the non-additive kinetic contributions were
computed using the revAPBEK kinetic functional \cite{apbe,apbek}
and a supermolecular def2-TZVPPD basis set \cite{def2tzvpp,furchepol}
was employed.
{As the aim of this work is not to verify the absolute accuracy of the
embedding approach (which depends critically on the KE approximation), but if
the additional errors due to the non-additive XC approximation can be reduced or
not, we believe that checking one (accurate) KE functional is enough.}

All calculations have been performed using the \texttt{FDE} 
script \cite{fde_hybrid} 
of the \texttt{TURBOMOLE} program package \cite{turbomole}.
{The calculation of the matrix elements of the non-additive TPSS-L XC functional (which is a Laplacian-level meta-GGA functional),  
has been performed as described in the Appendix A of Ref. \cite{fde_lap}.
}
  
The complexes considered for the tests have been divided into
four groups according to the character dominating their interaction:
\begin{itemize}
\item[-] WI (weak interaction): He-Ne, He-Ar, Ne-Ne, Ne-Ar, CH$_4$-Ne, C$_6$H$_6$-Ne, CH$_4$-CH$_4$; 
\item[-] DI (dipole-dipole interaction):  H$_2$S-H$_2$S, HCl-HCl, H$_2$S-HCl, CH$_3$Cl-HCl, CH$_3$SH-HCN, CH$_3$SH-HCl; 
\item[-] HB (hydrogen bond):  NH$_3$-NH$_3$, HF-HF, H$_2$O-H$_2$O, HF-HCN, (HCONH$_2$)$_2$, (HCOOH)$_2$;
\item[-] CT (charge transfer):  NF$_3$-HCN,C$_2$H$_4$-F$_2$,NF$_3$-HCN,
C$_2$H$_4$-Cl$_2$, NH$_3$-F$_2$, NH$_3$-ClF, NF$_3$-HF, C$_2$H$_2$-ClF,
HCN-ClF, NH$_3$-Cl$_2$, H$_2$O-ClF, NH$_3$-ClF.
\end{itemize}
The reference geometries and binding energies were
taken from Refs. \onlinecite{truhlar05a,truhlar05nb,wesolowski96fhnch,fde_ct}.

The error on the total embedding energy was computed as 
the difference between the energy obtained from 
a subsystem DFT calculation (i.e. Eq. (\ref{eq:ener})) 
and the energy ($E^\text{GKS}$) obtained from the
corresponding supermolecular conventional calculation \cite{fde_hybrid,apbek,fde_hyb_ene}, i.e.
\begin{equation}
\Delta E = E_{A+B}[\tilde{\rho}_A,\tilde{\rho}_B]-E^\text{GKS}[\rho_\text{GKS}]
\end{equation}
where $\tilde{\rho}_A$ and  $\tilde{\rho}_B$ are approximated embedded subsystems densities, due to the approximations in Eq. (\ref{eq:appfsnadd}).

The performance of the different approaches was evaluated, 
{\it within each group of complexes}, by computing the mean absolute
error (MAE)
and the mean absolute relative error (MARE)
with respect to reference binding energies \cite{fde_hyb_ene}. 
Instead, to assess the performance of the methods {\it for all the different classes 
of systems}, we considered the quantities \cite{apbek}
\begin{eqnarray}
\text{rwMAE}&=&\frac{1}{4} \sum_{i=\text{WI,DI,HB,CT}} \left( \frac{MAE_i}{\langle MAE_i\rangle}\right) \\
\text{rwMARE}&=&\frac{1}{4} \sum_{i=\text{WI,DI,HB,CT}} \left( \frac{MARE_i}{\langle MARE_i\rangle}\right)
\end{eqnarray}
where $\langle MAE_i\rangle$  ($\langle MARE_i\rangle$)  is the average MAE (MARE) among the different methods considered for the class of systems $i$.
In this way, all the different classes of systems will have the same influence of the rwMAE (rwMARE) and 
methods with rwMAE (rwMARE)  smaller than 1 will have better performance than the average.

The errors on the embedding densities were studied
by considering the deformation density
\begin{equation}
\Delta \rho(\mathbf{r}) = \tilde{\rho}_A(\mathbf{r}) + \tilde{\rho}_B(\mathbf{r}) - \rho^\text{GKS}(\mathbf{r})\ ,
\end{equation}
where $\rho^\text{GKS}$ denotes the density obtained from a conventional GKS calculation.
A quantitative measurement of the absolute error associated with a given 
embedding density was then obtained by computing the embedding density error
\begin{equation}\label{xi}
\xi=\frac{1000}{N}\int\left|
\Delta \rho(\R) \right|\,d \mathbf{r},
\end{equation}
with $N$ the number of electrons. In the evaluation of 
$\xi$ only
valence electron densities were considered.
Core densities are in fact much higher than valence ones and would
largely dominate.
On the other hand, core densities are not very important for the 
determination of chemical and physical properties of the interaction 
between the subsystems, which are of interest here.
The performance of the different approaches was evaluated by computing the MAE and the rwMAE.

\section{Results}\label{sec:res}
\begin{table*}
\begin{center}
\caption{\label{tab_dens}Errors on the embedding densities for different methods as obtained using Eq. (\ref{xi}).
 At the bottom of each group of results the mean absolute error (MAE) is reported. The last row report the global rwMAE (see text for details).
The best result of each line is highlighted in bold style.  A star indicates the best result among meta-GGA methods.}
\begin{ruledtabular}
\begin{tabular}{lrrrrr}
Complex  &  PBE/PBE  &  PBE0/PBE  &  TPSS/PBE  &  TPSS/TPSS-1  &  TPSS/TPSS-L  \\
\hline
\multicolumn{6}{c}{weak interaction (WI)} \\ 
He-Ne  &     0.05  &  \textbf{0.02}  &  0.05*  &  0.05*           &  0.05*  \\ 
He-Ar  &     \textbf{0.06}  &  \textbf{0.06}  &  0.12  &  0.08  &  0.05*  \\ 
Ne-Ne  &     0.04  &  \textbf{0.02}  &  0.06  &  0.04           &  0.03*  \\ 
Ne-Ar  &     0.06  &  \textbf{0.04}  &  0.13  &  0.08           &  0.05*  \\ 
CH$_4$-Ne  &    0.07  &  \textbf{0.05}  &  0.16  &  0.13       &  0.06*  \\ 
C$_6$H$_6$-Ne  &  0.13  &  \textbf{0.11}  &  0.25  &  0.27  &  0.13*  \\ 
CH$_4$-CH$_4$  &  0.60  &  \textbf{0.50}  &  0.84  &  0.79  &  0.53*  \\ 
MAE  &           0.14  &  \textbf{0.11}  &  0.23  &  0.21  &  0.13*  \\ 
\multicolumn{6}{c}{dipole-dipole interaction (DI)} \\ 
H$_2$S-H$_2$S  &  1.85  &  \textbf{1.62}  &  1.89  &  1.81  &  1.70*  \\ 
HCl-HCl  &  1.87  &  \textbf{1.49}  &  1.88  &  1.91  &  1.75  \\ 
H$_2$S-HCl  &  3.70  &  \textbf{2.97}  &  3.59  &  3.74  &  3.56  \\ 
CH$_3$Cl-HCl  &  2.38  &  \textbf{1.91}  &  2.36  &  2.40  &  2.24  \\ 
CH$_3$SH-HCN  &  1.72  &  \textbf{1.61}  &  1.74  &  1.64  &  1.58  \\ 
CH$_3$SH-HCl  &  4.08  &  \textbf{3.32}  &  3.90  &  4.12  &  3.95  \\ 
MAE  &  2.60  &  \textbf{2.15}  &  2.56  &  2.60  &  2.46*  \\ 
\multicolumn{6}{c}{hydrogen bond (HB)} \\ 
NH$_3$-NH$_3$  &  1.79  &  \textbf{1.67}  &  1.87  &  1.85  &  1.74  \\ 
HF-HF  &  1.53  &  \textbf{1.19}  &  1.56  &  1.62          &  1.50*  \\ 
H$_2$O-H$_2$O  &  2.01  &  \textbf{1.72}  &  2.05  &  2.11  &  1.98*  \\ 
NH$_3$-H$_2$O  &  3.11  &  \textbf{2.69}  &  3.06*  &  3.19  &  3.08  \\ 
HF-HCN  &  2.77  & \textbf{2.38}  &  2.62*  &  2.84          &  2.75  \\ 
(HCONH$_2$)$_2$  &  2.72  &  \textbf{2.49}  &  2.71*  &  2.76  &  2.65  \\ 
(HCOOH)$_2$  &  3.35  &  \textbf{2.94}  &  3.23*  &  3.45     &  3.37  \\ 
MAE  &  2.47  &  \textbf{2.15}  &  2.44*  &  2.55          &  2.54  \\ 
\multicolumn{6}{c}{charge transfer (CT)} \\ 
NF$_3$-HCN & 0.29 & \textbf{0.24} & 0.40 & 0.43 & 0.26* \\
C$_2$H$_4$-F$_2$  &  6.35  &  \textbf{2.75}  &  5.68*  &  5.79  &  5.79  \\ 
NF$_3$-HNC & 0.58 & \textbf{0.49} & 0.58 & 0.63 & 0.55* \\
C$_2$H$_4$-Cl$_2$ & 5.77 & \textbf{4.32} & 5.85* & 6.08 & 6.28 \\
NH$_3$-F$_2$  &  9.60  &  \textbf{4.38}  &  8.48  &  8.60  &  8.58*  \\ 
NF$_3$-ClF & 1.73 & \textbf{0.99} & 1.59 & 1.54* & 1.68 \\
NF$_3$-HF & 0.95 & \textbf{0.75} & 0.91 & 0.88* & 0.91 \\
C$_2$H$_2$-ClF  &  6.02  &  \textbf{4.32}  &  5.97*  &  6.77  &  6.51  \\ 
HCN-ClF  &  3.21  &  \textbf{2.33}  & 3.08*  &  3.40  &  3.30  \\ 
NH$_3$-Cl$_2$  &  7.60  &  \textbf{5.48}  &  7.42*  &  8.25  &  8.06  \\ 
H$_2$O-ClF  &  5.06  &  \textbf{3.42}  &  4.98  &  5.54  &  5.39*  \\ 
NH$_3$-ClF  &  11.19  &  \textbf{9.37}  &  11.00*  &  12.37  &  12.06  \\ 
MAE  &  4.86  &  \textbf{3.24}  &  4.66*  &  5.02  &  4.95  \\ 
  &    &    &    &    &    \\ 
\hline
rwMAE  &  1.00  &  \textbf{0.79}  &  1.12  &  1.14  &  0.97*  \\
\end{tabular}
\end{ruledtabular}
\end{center}
\end{table*}

The errors $\xi$ on the embedding densities for the different
methods are reported in Table \ref{tab_dens}.
The best performance is observed for the 
PBE0/PBE method, which gives the smallest error for
all the systems investigated.
As explained in Refs. \onlinecite{fde_hybrid,fde_ct,fde_lhf} 
this fact traces back to the reduced self-interaction
error of the PBE0 functional, which reduces the
overlap between the subsystem densities.
All other methods yield very close results with a rwMAE 
in the range $0.97-1.14$. 
Actually the TPSS/TPSS-L has the lowest rwMAE between these ones, 
showing that  
meta-GGA subsystem calculations can perform even better than 
conventional GGA calculations, despite the former include an additional
approximation.
Among meta-GGA methods, the TPSS/TPSS-L approach is 
the best 
for WI systems (MAE=0.13) and for DI (MAE=2.46), 
whereas TPSS/PBE is the best for HB and CT.

The integrated measure $\xi$ however provides only an absolute measure of
the error on embedding densities, but cannot tell anything 
on how the error on the density is distributed in the space and
what are the roles of the kinetic and XC approximations to determine such
an error.
Here, we aim at understanding better the importance
of different approximations used in the non-additive XC term
of the embedding potential.
\begin{figure}
\includegraphics[width=\columnwidth]{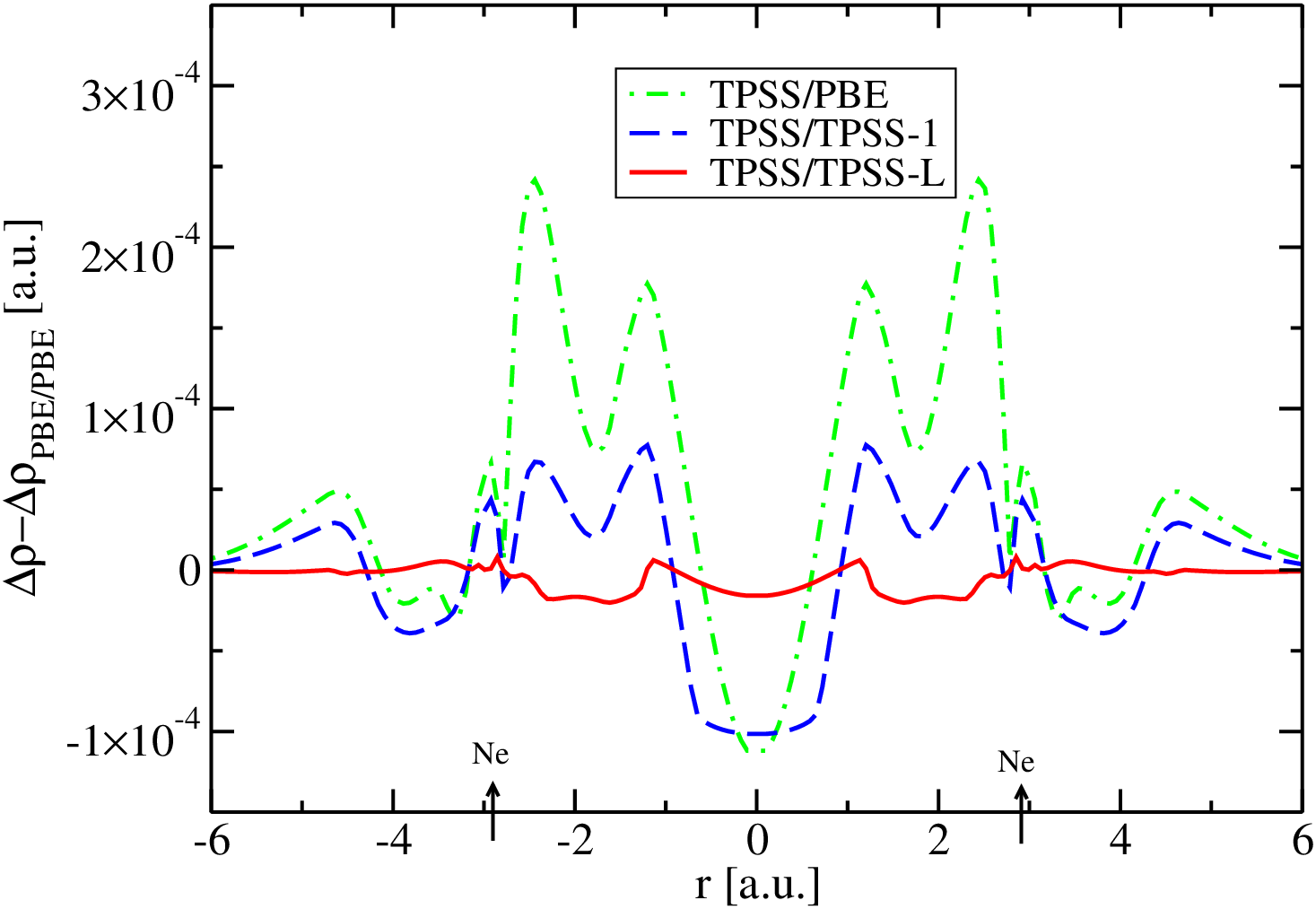}
\caption{\label{dendiff_fig} 
Plot of $\Delta\rho^{method}(z)-\Delta\rho_{PBE/PBE}(z)$ (for the Ne dimer along the main axis) 
computed using the kinetic energy density models of Eqs. (\ref{tau1}) and (\ref{tau2}) as well as the PBE XC functional for the non-additive XC embedding potential.}
\end{figure}
Thus, we consider in Fig. \ref{dendiff_fig}
the plot along the bond axis of the Ne-Ne complex (taken as an example)
of the plane-averaged XC deformation density
$\Delta\rho^{method}(z)-\Delta\rho^{PBE/PBE}(z)$,
where the plane-averaged deformation density is
\begin{equation}
\Delta\rho(z) = \int  | \Delta\rho(\mathbf{r})| dx dy\ .
\end{equation}  
This quantity provides in fact a measure, point by point, of the
embedding density error due to the XC approximation: the PBE/PBE is in fact taken as reference because it includes only approximation due to KE.
We see that, in agreement with the results of 
Fig. \ref{nenediff}, the TPSS-L approximation
performs very well, introducing only very small errors
in the calculation of the embedding density.
A larger effect is instead obtained for TPSS-1, while
the use of the PBE XC functional to compute the embedding
potential yields considerably larger differences.

\begin{table*}
\begin{center}
\caption{\label{tab_ene} Embedding energy errors (mHa) for different methods and complexes. 
Accurate reference binding energies ($E_{b}$) from Refs. \onlinecite{truhlar05a,truhlar05nb,wesolowski96fhnch} 
are also reported in the second column. At the bottom of each group of results the mean absolute error (MAE) and the 
mean absolute relative error (MARE) are reported. 
The best result of each line is highlighted in bold style. A star indicates the best result among the ones with the TPSS functional.
}
\begin{ruledtabular}
\begin{tabular}{lrrrrrr}
Complex  &  $E_b$  &  PBE/PBE  &  PBE0/PBE  &  TPSS/PBE  &  TPSS/TPSS-1  &  TPSS/TPSS-L  \\
\hline 
\multicolumn{7}{c}{weak interaction (WI) } \\ 
He-Ne         &  0.06  &  0.08    &  \textbf{0.03}  &  \textbf{0.03}*  &  0.06  &  0.08  \\ 
He-Ar         &  0.10  &  0.05    &  \textbf{0.00}  &  -0.01*          &  0.04  &       0.06  \\ 
Ne-Ne         &  0.13  &  0.14    &  0.06           &  \textbf{0.02}*  &  0.10  &  0.13  \\ 
Ne-Ar         &  0.21  &  0.11    & \textbf{0.04}   &  \textbf{-0.04}* &  0.06  &  0.11  \\ 
CH$_4$-Ne     &  0.35  &  0.12    &  \textbf{0.04}  &  \textbf{-0.04}*  &  0.06  &  0.12  \\ 
C$_6$H$_6$-Ne  &  0.75  &  -0.03  &  -0.10         &  -0.51            &  -0.25  &  \textbf{-0.01}*  \\ 
CH$_4$-CH$_4$  &  0.81  &  -0.38  &  -0.41         &  -0.82            &  -0.54  &  \textbf{-0.27}*  \\ 
MAE  &                 &  0.13  &  \textbf{0.10}  &  0.21  &  0.16  &  0.11*  \\ 
MARE  &               &  61\%  &   \textbf{27\%}  &  39\%*  &  52\%  &  59\%  \\ 
\multicolumn{7}{c}{dipole-dipole (DI) } \\ 
H$_2$S-H$_2$S  &  2.63  &  \textbf{-0.47}  &  -0.84  &  -1.16  &  -1.07  &  -0.49*  \\ 
HCl-HCl       &  3.20  &  0.07         &  -0.37      &  -0.70  &  -0.62  &  \textbf{-0.02}*  \\ 
H$_2$S-HCl    &  5.34  &  0.40         &  -0.42       &  -0.54  &  -0.71  &  \textbf{0.29}*  \\ 
CH$_3$Cl-HCl  &  5.66  &  \textbf{0.02}  &  -0.59     &  -1.14  &  -1.27  &  -0.05*  \\ 
CH$_3$SH-HCN  &  5.72  &  -1.18       &  -1.57        &  -2.09  &  -2.04  &  \textbf{-1.02}*  \\ 
CH$_3$SH-HCl  &  6.63  &  0.73        &  \textbf{-0.34}  &  -0.64  &  -1.06  &  0.54*  \\ 
MAE  &    &  0.48  &  0.69  &  1.05  &  1.13  &  \textbf{0.40}*  \\ 
MARE  &    & 10\%  &  16\%  &  24\%  &  25\%  &  \textbf{9\%}*  \\ 
\multicolumn{7}{c}{hydrogen bond (HB) } \\ 
NH$_3$-NH$_3$  &  5.02  &        -0.95  &  -1.32  &  -1.69  &  -1.63  &  \textbf{-0.80}*  \\ 
HF-HF         &  7.28  &  0.79            &  0.19  &  -0.13  &  \textbf{-0.03}*  &  0.78  \\ 
H$_2$O-H$_2$O  &  7.92  &  -0.20             &  -0.79  &  -1.11  &  -1.15  &  \textbf{-0.15}*  \\ 
NH$_3$-H$_2$O  &  10.21  &  -0.44           &  -1.28  &  -1.47  &  -1.75  &  \textbf{-0.36}*  \\ 
HF-HCN          &  11.33  &  \textbf{0.43} &  -0.56  &  -0.72  &  -1.06  &         0.49*   \\ 
(HCONH$_2$)$_2$  &  23.81  &  -4.21        &  -5.30  &  -5.95  &  -6.87  &  \textbf{-3.42}*  \\ 
(HCOOH)$_2$     &  25.74  &  -1.88         & -3.69  &  -3.94  &  -5.61  &  \textbf{-1.37}*  \\ 
MAE  &    &   1.27  &  1.88  &  2.14  &  2.59  &  \textbf{1.05}*  \\ 
MARE  &    &  9\%  &  13\%  &  16\%  &  18\%  &  \textbf{8\%}*  \\ 
\multicolumn{7}{c}{charge transfer (CT) } \\ 
NF$_3$-HCN        &	1.67 & -0.41  & -0.43        & -0.95 & -0.88 & \textbf{-0.31}* \\
C$_2$H$_4$-F$_2$  &  1.69   &  4.27  &  \textbf{1.92}  &  3.13*  &  3.42  & 3.87  \\ 
NF$_3$-HNC      & 2.31      & -0.13 & -0.51            &	-0.78 &	-1.11 &	\textbf{-0.02}* \\
C$_2$H$_4$-Cl$_2$ & 2.60   & 1.52     & \textbf{-0.42} & 0.30* & -1.87 & 1.70 \\
NH$_3$-F$_2$     &  2.88  &  6.90  &  \textbf{2.98}  &  5.17*  &  5.47  &  6.07  \\ 
NF$_3$-ClF     & 2.92    & 2.14   & 0.82 & 0.88 & \textbf{0.15}* & 1.95 \\
NF$_3$-HF      & 2.92    & 0.91   &  0.22 & \textbf{0.05}* & -0.57 & 0.86 \\
C$_2$H$_2$-ClF  &  6.07  &  3.71  &  \textbf{1.52}  &  2.40  &  1.64*  &  3.77  \\ 
HCN-ClF        &  7.74  &  1.62  &  \textbf{0.03}  &  0.28  &  -0.27*  &  1.51  \\ 
NH$_3$-Cl$_2$  &  7.78  &  2.84  &  \textbf{0.21}  &  1.64  &  0.85*  &  2.94  \\ 
H$_2$O-ClF     &  8.54  &  2.42  &  \textbf{0.45}  &  1.17  &  0.55*  &  2.45  \\ 
NH$_3$-ClF     &  16.92  &  4.44  &  1.31  &  2.35  &  \textbf{-0.33}*  &  5.75  \\ 
MAE  &    &  2.61  &  \textbf{0.90}  &  1.59  &  1.43*  &  2.60  \\ 
MARE  &    &  72\%  & \textbf{30\%}  &  49\%*  &  53\%  &  67\%  \\ 
  &    &    &    &    &    &    \\ 
\hline
rwMAE  &&  0.93 & \textbf{0.79}  	& 1.24	& 1.21	& 0.84* \\
rwMARE &&  0.98\% &  \textbf{0.78}\%	 & 1.10\%& 1.24\% & 0.91\%* \\
\end{tabular}
\end{ruledtabular}
\end{center}
\end{table*}
We now turn to discuss the embedding energy errors, which are reported in Tab. \ref{tab_ene}.
In this case the hybrid PBE0/PBE is not the best for all systems, 
as it was in Tab. \ref{tab_dens}.
In fact, for dipole-dipole and hydrogen bond complexes 
the TPSS/TPSS-L method is the most accurate, closely followed
by the PBE/PBE approach, whereas the PBE0/PBE is not so accurate 
for these systems \cite{fde_hyb_ene,fde_lhf}.
On the other hand, PBE0/PBE is very accurate for 
weakly-interacting and charge-transfer systems \cite{fde_ct}.

A more detailed discussion of the trends is provided in
subsection \ref{sec:decomp} 
where we perform an energy decomposition analysis of 
the embedding energy errors.
Here we just note that, concerning the meta-GGA approaches,
our simple Laplacian-level meta-GGA approximation
(TPSS/TPSS-L) is significantly more accurate than the GGA
TPSS/TPSS-1 and TPSS/PBE ones and can also slightly outperform
the use of a simple GGA XC functional, such as PBE.
In fact, for TPSS/TPSS-L, PBE/PBE, TPSS/PBE, and TPSS/TPSS-1
we find overall rwMAEs of 0.84, 0.93 and 1.21, 1.24, 
respectively.

\subsection{Error decomposition analysis}
\label{sec:decomp}
The embedding energy error for meta-GGA as well as for hybrid functionals 
depends on two distinct approximations, the KE and the XC energy. As discussed in Ref. \onlinecite{fde_hyb_ene} 
this causes a subtle error cancellation effect.
To understand better the error compensation issue
in meta-GGA subsystem DFT calculations
and analyze in detail the sources of different
errors, we perform in the following
an error decomposition analysis.

Following Ref. \onlinecite{fde_hyb_ene} we thus write
the error on the embedding energy as
\begin{equation}
\Delta E = \Delta T_s + \Delta D + 
\Delta E_{xc}\ ,
\end{equation}
with
\begin{eqnarray}
\Delta T_s & = & \widetilde{T}_s^{nadd}[\tilde{\rho}_A,\tilde{\rho}_B] - \\
\nonumber
&& -\left(T_s[\{\phi^\text{GKS}_{A+B}\}]-
T_s[\{\tilde{\phi}^A\}]-T_s[\{\tilde{\phi}^B\}]\right)\\
\Delta D & = & E_{ext}[\tilde{\rho}_{A+B}] + J[\tilde{\rho}_{A+B}] 
+ \tilde{E}_{xc}[\tilde{\rho}_{A+B}]
- \\
\nonumber
&& - \left(E_{ext}[\rho^\text{GKS}] + J[\rho^\text{GKS}] + \widetilde{E}_{xc}
[\rho^\text{GKS}]
\right)\\
\Delta E_{xc} & = & \widetilde{E}_{xc}[\rho^\text{GKS}] -
                    \widetilde{E}_{xc}[\tilde{\rho}_A] -
                    \widetilde{E}_{xc}[\tilde{\rho}_B] - \\
\nonumber
&&                   - \left(E_{xc}[\{\phi^\text{GKS}_{A+B}\}]-
                         E_{xc}[\{\tilde{\phi}^A\}]-
                         E_{xc}[\{\tilde{\phi}^B\}]\right)\ .
\end{eqnarray}
Here the first term ($\Delta T_s$) describes the error
associated with the KE approximation,
the second term ($\Delta D$) is a relaxation term
directly related to the embedding density error,
and the third one ($\Delta E_{xc}$) measures the
error due to the XC approximation.
This latter term will be negative when the approximate 
XC functional overestimates with respect to the
non-local one and positive in the opposite case.
{In analogy to the previous notation, 
$\tilde{\phi}_A$ and  $\tilde{\phi}_B$ denote 
single particle KS orbitals of subsystems $A$ and $B$, respectively,
as obtained from approximated embedding calculations.}

The results of the energy decomposition analysis are
reported in Table \ref{tab_ene_dec}.
The contributions due to $\Delta T_s$ and $\Delta D$ are reported summed 
together because both terms yield very large values but with opposite sign,
thus they only contribute to the total error through a strong error cancellation.

\begin{table*}
\begin{center}
\caption{\label{tab_ene_dec} Embedding energy error decomposition (mHa) for different methods and complexes. 
At the bottom of each group of results the mean absolute error (MAE) and the mean absolute relative error (MARE) are reported. 
In addition for $\Delta E_{xc}$ also the XC differential error (XCDE) and the XC differential relative error (XCDRE) are listed.}
\begin{ruledtabular}
\begin{footnotesize}
\begin{tabular}{lrrrrrr}
Complex  &  \multicolumn{2}{c}{TPSS/PBE}  & \multicolumn{2}{c}{TPSS/TPSS-1}  &  \multicolumn{2}{c}{TPSS/TPSS-L} \\
\cline{2-3}\cline{4-5}\cline{6-7}
 & $\Delta T_s+\Delta D$ & $\Delta E_{xc}$ & $\Delta T_s+\Delta D$ & $\Delta E_{xc}$ & $\Delta T_s+\Delta D$ & $\Delta E_{xc}$ \\ 
\hline
\multicolumn{7}{c}{weak interaction} \\
He-Ne  &  0.08  &  -0.05           &  0.08  &  -0.02  &  0.08  &  -0.01  \\ 
He-Ar  &  0.05  &  -0.06           &  0.05  &  -0.02  &  0.05  &   0.00  \\ 
Ne-Ne  &  0.13  &  -0.11           &  0.13  &  -0.04  &  0.14  &  -0.01  \\ 
Ne-Ar  &  0.10  &  -0.14           &  0.11  &  -0.05  &  0.12  &   0.00  \\ 
CH$_4$-Ne  &  0.10  &  -0.14       &  0.11  &  -0.05  &  0.11  &   0.00  \\ 
C$_6$H$_6$-Ne  &  -0.06  &  -0.46  &  -0.05  &  -0.20  &  -0.01  &  0.00  \\ 
CH$_4$-CH$_4$  &  -0.43  &  -0.39  &  -0.42  &  -0.12  &  -0.38  &  0.11  \\ 
MAE  &  0.14  &  0.19  &  0.14  &  0.07   &  0.13  &  0.02  \\ 
MARE  &  60\%  &  63\%  &  61\%  &  23\%  &  61\%  &  5\%  \\ 
XCDE/XCDRE  &    &  +0.08/-21\%  &    &  +0.02/-12\%  &        & -0.02/-5\%  \\ 
\multicolumn{7}{c}{dipole-dipole} \\
H$_2$S-H$_2$S  &  -0.48  &  -0.68  &  -0.45  &  -0.62  &  -0.28  &  -0.21  \\ 
HCl-HCl  &      0.06  &  -0.76     &  0.08  &  -0.71  &  0.19 &  -0.16  \\ 
H$_2$S-HCl  &  0.38  &  -0.93      &  0.44  &  -1.15  &  0.67  &  -0.38  \\ 
CH$_3$Cl-HCl  &  0.01  &  -1.15    &  0.02  &  -1.29  &  0.22  &  -0.28  \\ 
CH$_3$SH-HCN  &  -1.20  &  -0.90   &  -1.21  &  -0.84  &  -0.99  &  -0.03  \\ 
CH$_3$SH-HCl  &  0.71  &  -1.35    &  0.73  &  -1.80  &  1.20  &  -0.67  \\ 
MAE  &  0.47  &  0.96  &  0.49  &  1.07     &  0.59  &  0.29  \\ 
MARE  & 10\%  &  21\%  &  10\%  &  22\%    &  11\%  &  6\%  \\ 
XCDE/XCDRE  &    &  +0.58/+14\%      &    &  +0.65/+15\%        &      &  -0.19/-3\% \\ 
\multicolumn{7}{c}{hydrogen bond} \\
NH$_3$-NH$_3$  &  -0.96  &  -0.74  &  -0.96  &  -0.67  &  -0.88  &   0.08  \\ 
HF-HF        &  0.79  &  -0.92  &  0.79  &  -0.82       &  0.89  &   -0.11  \\ 
H$_2$O-H$_2$O  &  -0.21  &  -0.90  &  -0.20  &  -0.95  &  -0.02  &  -0.13  \\ 
NH$_3$-H$_2$O  &  -0.46  &  -1.01  &  -0.45  &  -1.30  &  -0.13  &  -0.23  \\ 
HF-HCN      &  0.41  &  -1.13  &  0.44  &  -1.51       &  0.68  &   -0.19  \\ 
(HCONH$_2$)$_2$  &  -4.20  &  -1.74  &  -4.36  &  -2.52  &  -3.49  &  0.07  \\ 
(HCOOH)$_2$  &  -1.87  &  -2.07     &  -2.04  &  -3.57  &  -1.31  &  -0.06 \\ 
MAE  &  1.27  &  1.22  &  1.32  &  1.62  &  1.06  & 0.12  \\ 
MARE  &  9\%  &  11\%  &  10\%  &  12\%  &  8\%  &  1\%  \\ 
XCDE/XCDRE  &       &  +0.87/+6\%  &        &  +1.27/+8\%   &      &  +0.00/+0\%  \\ 
\multicolumn{7}{c}{charge transfer} \\
NF$_3$-HCN  &  -0.46  &  -0.50  &  -0.44  &  -0.19      &  -0.36  &  0.04  \\ 
C$_2$H$_4$-F$_2$  &  4.23  &  -1.10  &  4.21  &  -0.79  &  3.83  &  0.04  \\ 
NF$_3$-HNC  &  -0.14  &  -0.63  &  -0.13  &  -0.49      &  0.00  &  -0.02  \\ 
C$_2$H$_4$-Cl$_2$  &  1.51  &  -1.21  &  1.55  &  -2.05  &  1.66  &  0.04  \\ 
NH$_3$-F$_2$  &  6.84  &  -1.67  &  6.94  &  -1.48      &  6.38  &  -0.31  \\ 
NF$_3$-ClF  &  2.11  &  -1.22  &  2.17  &  -1.25       &  2.11  &  -0.17  \\ 
NF$_3$-HF  &  0.89  &  -0.84  &  0.92  &  -0.83       &  1.03  &  -0.16  \\ 
C$_2$H$_2$-ClF  &  3.71  &  -1.31  &  3.84  &  -2.21  &  3.79  &  -0.02  \\ 
HCN-ClF  &  1.60  &  -1.32  &  1.68  &  -1.95         &  1.63  &  -0.12  \\ 
NH$_3$-Cl$_2$  &  2.84  &  -1.19  &  2.90  &  -2.05   &  3.07  &  -0.14  \\ 
H$_2$O-ClF  &  2.41  &  -1.24  &  2.50  &  -1.94       &  2.54  &  -0.09  \\ 
NH$_3$-ClF  &  4.46  &  -2.11  &  4.27  &  -4.60       &  5.88  &  -0.13  \\ 
MAE  &  2.60  &  1.20  &  2.63  &  1.65  &  2.69  &  0.11  \\ 
MARE  &  71\%  &  32\%  &  72\%  &  35\%  &  69\%  &  3\%  \\ 
XCDE/XCDRE  &    &  -1.01/-22\%  &    &  -1.36/-25\%  &            &  -0.09/-2\%  \\ 
\hline
MAE    &  1.37  & 0.94         &  1.40  &  1.19        & 1.38  &  0.13  \\ 
MARE   &  44\%  & 32\%         &  44\%  &  25\%        &  43\%  &  4\%  \\ 
XCDE/XCDRE &    &  -0.06/-9\%  &        &  -0.11/-7\%  &     &  -0.07/-3\%  \\ 
\end{tabular}
\end{footnotesize}
\end{ruledtabular}
\end{center}
\end{table*}
For each group of complexes as well as for the overall test set
we report, for each component of the energy decomposition, the MAE and the MARE. Moreover, for $\Delta E_{xc}$
we report the XC differential error (XCDE) and the XC 
differential relative error (XCDRE), defined as
\begin{eqnarray}
\mathrm{XCDE} & = & \frac{1}{N}\sum_{i=1}^N\left|\Delta E_i\right|-
\left|\Delta T_{s,i}+\Delta D_i \right|\\
\mathrm{XCDRE} & = & \frac{1}{N}\sum_{i=1}^N\frac{\left|\Delta E_i\right|-
\left|\Delta T_{s,i}+\Delta D_i\right|}{E_{b,i}}\ ,
\end{eqnarray}
where the sums extend over all the $N$ systems in the set.
A positive value of these statistical indicators denotes that the XC approximation 
has a bad effect on the absolute total embedding energy error, 
increasing it. On the contrary,
a negative value indicates that the employed XC approximation reduces the error 
(presumably by error cancellation).

Inspection of the table shows that the $\Delta T_s + \Delta D$ values are very
similar for all the 
methods.
This reflects the fact that we used the {\it same}
KE approximation in all calculations and that in all cases 
the final embedding densities are quite similar.
On the other hand, the values of $\Delta E_{xc}$ are more differentiated between the 
various methods. In particular, much smaller values are found in general
for the TPSS/TPSS-L method (global MARE 4\%).
The  TPSS/TPSS-1 (global MARE 25\%) and TPSS/PBE (global MARE 32\%) accuracy is much less.
This confirms that the TPSS/TPSS-L benefits of a much better XC approximation than
the latter.

Valuable information is also obtained by the inspection of
the XCDRE indicators.
For WI and CT systems the TPSS/PBE and TPSS/TPSS-1 methods
have negative values of XCDRE. Thus the additional error due to the XC approximation yields
(due to error cancellation) better total energies. This explains the results in Tab. \ref{tab_ene}, where  
TPSS/PBE (and TPSS/TPSS-1) shows a good accuracy for these systems.
On the other hand, for DI and HB systems, the XCDRE values are positive, i.e. the additional error due to the 
XC approximation reduces the accuracy of the embedding energy. In these cases no error cancellation
occurs and in fact TPSS/PBE and TPSS/TPSS-1 yield quite bad total energy (see Tab. \ref{tab_ene}).

On the other hand, in the TPSS/TPSS-L method the XCDRE values are very small (and negative), showing 
the smallest error compensation in relation to the  XC approximation.
Note, in addition, that for all the considered TPSS subsystem DFT calculations
the $\Delta E_{xc}$ values are comparable or smaller than for the hybrid
PBE0/PBE method (see Table II of Ref. \onlinecite{fde_hyb_ene} and note
that these values include a prefactor 0.25).


{Finally, we note that the good accuracy of the TPSS/TPSS-L approach is maintained also for larger subsystems' density
overlaps. This is shown in Tab. \ref{tabnoneq}, where we report the 
$\Delta E_{xc}$, $\Delta T_s + \Delta D$, and the density error $\xi$, computed with the TPSS/TPSS-L approach 
for several complexes at different
intermolecular distances (smaller distances correspond to higher overlaps).
Tab.  \ref{tabnoneq} also reports the XC absolute ratio (XCAR), defined as  
$XCAR=|\Delta E_{xc}|/(|\Delta T_s + \Delta D|+|\Delta E_{xc}|)$, which is an absolute (i.e. without error compensation) 
measure of the non-additive XC contribution to the total embedding energy error.
The data reported in Table   \ref{tabnoneq} clearly show that at shortest distance 
XCAR is not largely increased (as it happens instead for $\xi$), 
but remains constant and in some cases it is even reduced.  
}
\begin{table}
\begin{center}
\caption{\label{tabnoneq} Components of the embedding energy error decomposition 
($\Delta E_{xc}$ and $\Delta T_s + \Delta D$; mHa), XCAR (see text for details),   
and embedding density errors ($\xi$), from  TPSS/TPSS-L calculations (last column report  in parenthesis the value of 
$\xi$ for  PBE/PBE)
for various test complexes as a function of the intermolecular distance. $R_0$ denotes the
equilibrium distance in the complexes (Ne$_2$, $R_0=3.091$\AA; HF-NCH,
$R_0=1.805$\AA; (HCl)$_2$, $R_0=3.872$\AA).
}
\begin{tabular}{llrrrc}
\hline\hline
System & $R/R_0$ & $\Delta E_{xc}$ & $\Delta T_s + \Delta D$ & XCAR & $\xi$  \\
\hline
Ne$_2$    & 0.8 &  0.02 &  -0.17  & 10\%  & 0.34 (0.34) \\
          & 1.0 & -0.01 &   0.14  &  6\%  & 0.03 (0.04) \\ 
          & 1.5 &  0.00 &   0.00  &  0\%  & 0.00 (0.00) \\
(HCl)$_2$ & 0.8 & -1.55 &  -3.40  &  31\% & 9.31 (9.82) \\
          & 1.0 & -0.16 &   0.19  &  45\% & 1.75 (1.87) \\ 
          & 1.5 &  0.00 &   0.03  &  3\%  & 0.05 (0.05) \\
HF-NCH    & 0.8 & -0.58 &  -1.81  &  24\% & 5.49 (5.48) \\
          & 1.0 & -0.19 &   0.68  &  22\% & 2.75 (2.77) \\ 
          & 1.5 & -0.02 &   0.39  &  5\%  & 0.32 (0.37) \\
\hline\hline
\end{tabular}
\end{center}
\end{table}

\section{Conclusions}
\label{concl}
Using the generalized Kohn-Sham framework, 
we extended the subsystem DFT formalism to the use
of meta-GGA functionals. For a practical application of the
method we proposed several semilocal approximations for the
non-additive XC energy. Two of these are based on
simple models for the KED.

The results of our subsystem DFT calculations show
that all the proposed methods work reasonably well,
giving density and energy embedding errors 
comparable with conventional calculations
and close to hybrid subsystem DFT results.
Nevertheless, a more detailed analysis shows
that for the simplest approaches the final
performance is the result of a 
quite significant error compensation.
This effect is reduced only when more
sophisticated approximations for the non-additive
XC term are used.
In this respect we showed that this goal
can be pursued by considering Laplacian-level
meta-GGA approximations. Anyway,
we remark that our TPSS-L approximation,
despite giving promising results, is
only a simple model used here to
investigate the power of Laplacian-dependent
approximations and more work should be done on this topic.

In summary, we see several new research directions
that can be opened by the present work.
Firstly, subsystem DFT applications can surely benefit from
the use of meta-GGA functionals which provide increased accuracy
with respect to GGAs at a lower computational cost
than hybrid methods.
In this context new meta-GGA XC functionals can be tested
apart from TPSS.
In second place, additional research can be done to
develop more accurate semilocal approximations
for the KED, which can be useful
in the calculation of the non-additive XC energy.

Finally, additional work can be foreseen to
exploit the full power of the Laplacian level of theory,
investigating the use of Laplacian-level meta-GGA kinetic
energy functionals \cite{fde_lap} in conjunction with
similar approximations employed in the
non-additive XC term. 
In this way density and embedding errors 
in the kinetic and XC part can be expected to be more balanced.

\textbf{Acknowledgements}
This work was partially supported by the National
Science Center under Grant No.~DEC-2012/05/N/ST4/02079, and
No.~DEC-2013/08/T/ST4/00032.
We thank TURBOMOLE GmbH for providing the TURBOMOLE program package.

\appendix

\begin{figure}
\includegraphics[width=\columnwidth]{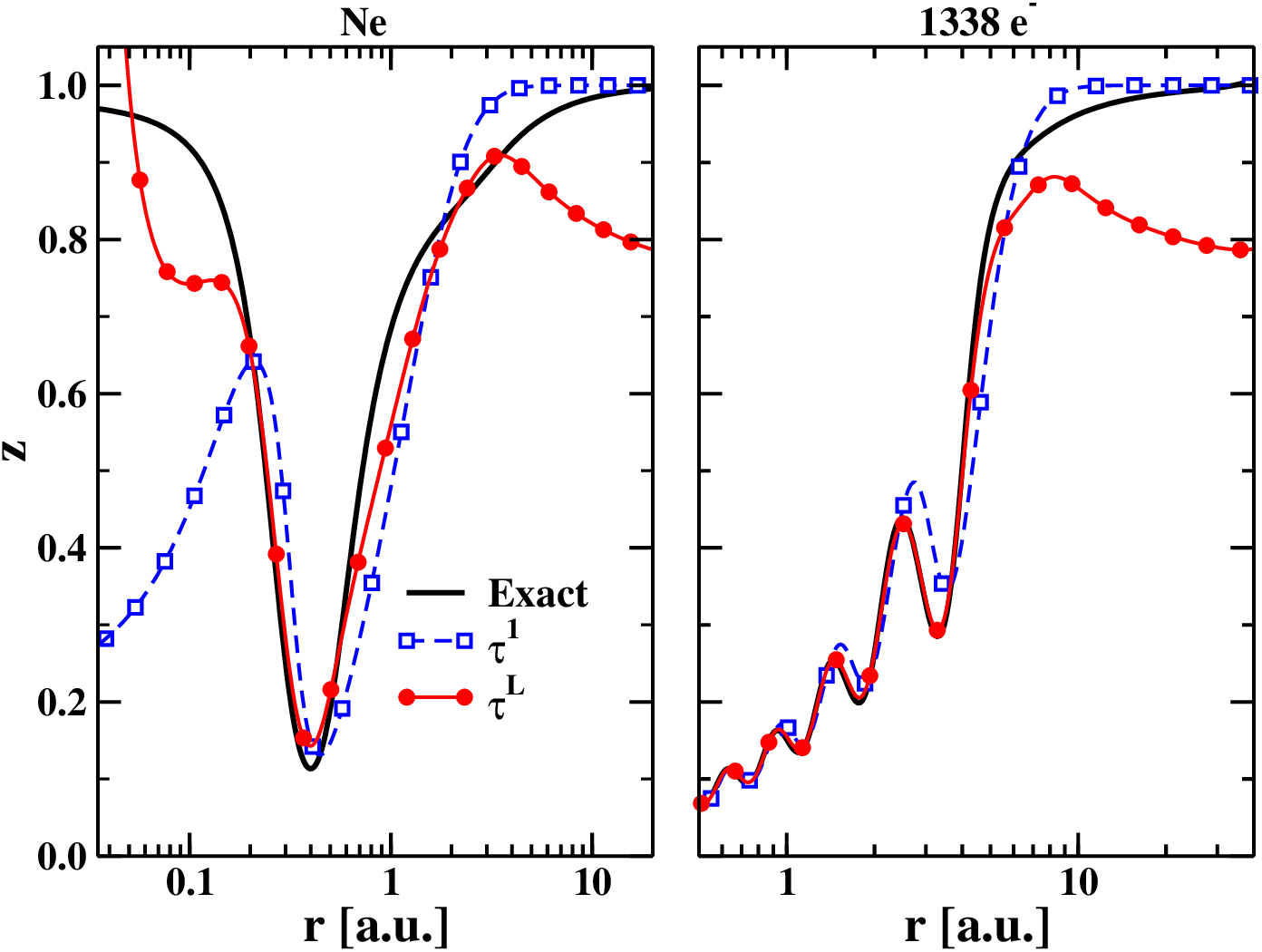}
\caption{\label{zfig} 
Plot of $z=\tau^\text{W}/\tau$ for the Ne atom and a non relativistic atom
with 1338 electrons, as computed with the exact $\tau$, $\tau=\tau^1$, 
and $\tau=\tau^L$.}
\end{figure}

\section{Behavior of $\tau^{revAPBEk} + (20/9)\tau^{TF}q$
in the tail of the density}
\label{appa}

Let consider a spherical atom, where the density decays exponentially as $\rho=A e^{-\alpha r}$, with 
$r$ being the radial distance from the nucleus. In this case, 
\begin{equation}
\frac{s^2}{q}=\frac{\alpha r}{\alpha r -2} \xrightarrow[{r\rightarrow\infty}]{} 1.
\label{eq1}
\end{equation}
A similar expression can also be obtained for a 
Gaussian decaying density $n=A e^{-\alpha r^2}$
\begin{equation}
\frac{s^2}{q}=\frac{2 \alpha r^2}{2 \alpha r^2 -3} \xrightarrow[{r\rightarrow\infty}]{} 1.
\label{eq2}
\end{equation}
Thus, in the tail of the density $q\approx s^2$, and so 
\begin{equation}
\tau^{revAPBEk} + \frac{20}{9}\tau^{TF}q \xrightarrow[{r\rightarrow\infty}]{} \frac{20}{9}\tau^{TF}s^2,
\label{eq3}  
\end{equation}
where we considered that the revAPBEK enhancement 
factor asymptotically behaves as a constant.
Finally we find
\begin{equation}
z\xrightarrow[{r\rightarrow\infty}]{}\frac{5/3}{20/9}=\frac{3}{4}.
\label{eq4}
\end{equation}
Note, however, that this behavior is valid only at large distances 
that are not important in practical calculations, see Fig. \ref{zfig}.
In valence and close tail regions $z$ is instead very close to the exact
one, as shown in Fig. \ref{nenediff} and Fig. \ref{zfig}.


\begin{thebibliography}{113}
\expandafter\ifx\csname natexlab\endcsname\relax\def\natexlab#1{#1}\fi
\expandafter\ifx\csname bibnamefont\endcsname\relax
  \def\bibnamefont#1{#1}\fi
\expandafter\ifx\csname bibfnamefont\endcsname\relax
  \def\bibfnamefont#1{#1}\fi
\expandafter\ifx\csname citenamefont\endcsname\relax
  \def\citenamefont#1{#1}\fi
\expandafter\ifx\csname url\endcsname\relax
  \def\url#1{\texttt{#1}}\fi
\expandafter\ifx\csname urlprefix\endcsname\relax\def\urlprefix{URL }\fi
\providecommand{\bibinfo}[2]{#2}
\providecommand{\eprint}[2][]{\url{#2}}

\bibitem[{\citenamefont{Cortona}(1991)}]{cortona}
\bibinfo{author}{\bibfnamefont{P.}~\bibnamefont{Cortona}},
  \bibinfo{journal}{Phys. Rev. B} \textbf{\bibinfo{volume}{44}},
  \bibinfo{pages}{8454} (\bibinfo{year}{1991}).

\bibitem[{\citenamefont{Wesolowski and Weber}(1996)}]{wesoiter}
\bibinfo{author}{\bibfnamefont{T.~A.} \bibnamefont{Wesolowski}}
  \bibnamefont{and} \bibinfo{author}{\bibfnamefont{J.}~\bibnamefont{Weber}},
  \bibinfo{journal}{Chem. Phys. Lett.} \textbf{\bibinfo{volume}{248}},
  \bibinfo{pages}{71 } (\bibinfo{year}{1996}).

\bibitem[{\citenamefont{Wesolowski}(2006)}]{wesorev}
\bibinfo{author}{\bibfnamefont{T.~A.} \bibnamefont{Wesolowski}}, in
  \emph{\bibinfo{booktitle}{Chemistry: Reviews of Current Trends}}, edited by
  \bibinfo{editor}{\bibfnamefont{J.}~\bibnamefont{Leszczynski}}
  (\bibinfo{publisher}{World Scientific: Singapore, 2006},
  \bibinfo{address}{Singapore}, \bibinfo{year}{2006}),
  vol.~\bibinfo{volume}{10}, pp. \bibinfo{pages}{1--82}.

\bibitem[{\citenamefont{Neugebauer}(2010)}]{neugrev10}
\bibinfo{author}{\bibfnamefont{J.}~\bibnamefont{Neugebauer}},
  \bibinfo{journal}{Phys. Rep.} \textbf{\bibinfo{volume}{489}},
  \bibinfo{pages}{1} (\bibinfo{year}{2010}).

\bibitem[{\citenamefont{Jacob and Neugebauer}(2014)}]{subdftrev}
\bibinfo{author}{\bibfnamefont{C.~R.} \bibnamefont{Jacob}} \bibnamefont{and}
  \bibinfo{author}{\bibfnamefont{J.}~\bibnamefont{Neugebauer}},
  \bibinfo{journal}{Wiley Interdisciplinary Reviews: Computational Molecular
  Science} \textbf{\bibinfo{volume}{4}}, \bibinfo{pages}{325}
  (\bibinfo{year}{2014}).

\bibitem[{\citenamefont{Krishtal et~al.}(2015)\citenamefont{Krishtal, Sinha,
  Genova, and Pavanello}}]{pavanello15}
\bibinfo{author}{\bibfnamefont{A.}~\bibnamefont{Krishtal}},
  \bibinfo{author}{\bibfnamefont{D.}~\bibnamefont{Sinha}},
  \bibinfo{author}{\bibfnamefont{A.}~\bibnamefont{Genova}}, \bibnamefont{and}
  \bibinfo{author}{\bibfnamefont{M.}~\bibnamefont{Pavanello}},
  \bibinfo{journal}{J. Phys. Cond. Matt.}  (\bibinfo{year}{2015}),
  \bibinfo{note}{in press}.

\bibitem[{\citenamefont{Hohenberg and Kohn}(1964)}]{hk}
\bibinfo{author}{\bibfnamefont{P.}~\bibnamefont{Hohenberg}} \bibnamefont{and}
  \bibinfo{author}{\bibfnamefont{W.}~\bibnamefont{Kohn}},
  \bibinfo{journal}{Phys. Rev.} \textbf{\bibinfo{volume}{136}},
  \bibinfo{pages}{B864} (\bibinfo{year}{1964}).

\bibitem[{\citenamefont{Kohn and Sham}(1965)}]{ks}
\bibinfo{author}{\bibfnamefont{W.}~\bibnamefont{Kohn}} \bibnamefont{and}
  \bibinfo{author}{\bibfnamefont{L.~J.} \bibnamefont{Sham}},
  \bibinfo{journal}{Phys. Rev.} \textbf{\bibinfo{volume}{140}},
  \bibinfo{pages}{A1133} (\bibinfo{year}{1965}).

\bibitem[{\citenamefont{Wesolowski et~al.}(1996)\citenamefont{Wesolowski,
  Chermette, and Weber}}]{wesolowski96fhnch}
\bibinfo{author}{\bibfnamefont{T.~A.} \bibnamefont{Wesolowski}},
  \bibinfo{author}{\bibfnamefont{H.}~\bibnamefont{Chermette}},
  \bibnamefont{and} \bibinfo{author}{\bibfnamefont{J.}~\bibnamefont{Weber}},
  \bibinfo{journal}{J. Chem. Phys.} \textbf{\bibinfo{volume}{105}},
  \bibinfo{pages}{9182} (\bibinfo{year}{1996}).

\bibitem[{\citenamefont{Wesolowski}(1997)}]{weso97}
\bibinfo{author}{\bibfnamefont{T.~A.} \bibnamefont{Wesolowski}},
  \bibinfo{journal}{J. Chem. Phys.} \textbf{\bibinfo{volume}{106}},
  \bibinfo{pages}{8516} (\bibinfo{year}{1997}).

\bibitem[{\citenamefont{Wesołowski et~al.}(1998)\citenamefont{Wesołowski,
  Ellinger, and Weber}}]{Wesolowski98}
\bibinfo{author}{\bibfnamefont{T.~A.} \bibnamefont{Wesołowski}},
  \bibinfo{author}{\bibfnamefont{Y.}~\bibnamefont{Ellinger}}, \bibnamefont{and}
  \bibinfo{author}{\bibfnamefont{J.}~\bibnamefont{Weber}}, \bibinfo{journal}{J.
  Chem. Phys.} \textbf{\bibinfo{volume}{108}} (\bibinfo{year}{1998}).

\bibitem[{\citenamefont{Wesolowski and Tran}(2003)}]{wesotran03}
\bibinfo{author}{\bibfnamefont{T.~A.} \bibnamefont{Wesolowski}}
  \bibnamefont{and} \bibinfo{author}{\bibfnamefont{F.}~\bibnamefont{Tran}},
  \bibinfo{journal}{J. Chem. Phys.} \textbf{\bibinfo{volume}{118}},
  \bibinfo{pages}{2072} (\bibinfo{year}{2003}).

\bibitem[{\citenamefont{Kevorkyants et~al.}(2006)\citenamefont{Kevorkyants,
  Dulak, and Wesolowski}}]{kevorkyants06hyd}
\bibinfo{author}{\bibfnamefont{R.}~\bibnamefont{Kevorkyants}},
  \bibinfo{author}{\bibfnamefont{M.}~\bibnamefont{Dulak}}, \bibnamefont{and}
  \bibinfo{author}{\bibfnamefont{T.~A.} \bibnamefont{Wesolowski}},
  \bibinfo{journal}{J. Chem. Phys.} \textbf{\bibinfo{volume}{124}},
  \bibinfo{eid}{024104} (\bibinfo{year}{2006}).

\bibitem[{\citenamefont{Dulak and Wesolowski}(2007)}]{dulak07}
\bibinfo{author}{\bibfnamefont{M.}~\bibnamefont{Dulak}} \bibnamefont{and}
  \bibinfo{author}{\bibfnamefont{T.~A.} \bibnamefont{Wesolowski}},
  \bibinfo{journal}{J. Molec. Model.} \textbf{\bibinfo{volume}{13}},
  \bibinfo{pages}{631} (\bibinfo{year}{2007}).

\bibitem[{\citenamefont{Dułak et~al.}(2007)\citenamefont{Dułak, Kamiński,
  and Wesołowski}}]{dulak07geom}
\bibinfo{author}{\bibfnamefont{M.}~\bibnamefont{Dułak}},
  \bibinfo{author}{\bibfnamefont{J.~W.} \bibnamefont{Kamiński}},
  \bibnamefont{and} \bibinfo{author}{\bibfnamefont{T.~A.}
  \bibnamefont{Wesołowski}}, \bibinfo{journal}{J. Chem. Theory Comput.}
  \textbf{\bibinfo{volume}{3}}, \bibinfo{pages}{735} (\bibinfo{year}{2007}).

\bibitem[{\citenamefont{{Garcia Lastra} et~al.}(2008)\citenamefont{{Garcia
  Lastra}, Kaminski, and Wesolowski}}]{lastra08}
\bibinfo{author}{\bibfnamefont{J.~M.} \bibnamefont{{Garcia Lastra}}},
  \bibinfo{author}{\bibfnamefont{J.~W.} \bibnamefont{Kaminski}},
  \bibnamefont{and} \bibinfo{author}{\bibfnamefont{T.~A.}
  \bibnamefont{Wesolowski}}, \bibinfo{journal}{J. Chem. Phys.}
  \textbf{\bibinfo{volume}{129}}, \bibinfo{eid}{074107} (\bibinfo{year}{2008}).

\bibitem[{\citenamefont{G\"{o}tz et~al.}(2009)\citenamefont{G\"{o}tz, Beyhan,
  and Visscher}}]{gotz09}
\bibinfo{author}{\bibfnamefont{A.~W.} \bibnamefont{G\"{o}tz}},
  \bibinfo{author}{\bibfnamefont{S.~M.} \bibnamefont{Beyhan}},
  \bibnamefont{and} \bibinfo{author}{\bibfnamefont{L.}~\bibnamefont{Visscher}},
  \bibinfo{journal}{J. Chem. Theory Comput.} \textbf{\bibinfo{volume}{5}},
  \bibinfo{pages}{3161} (\bibinfo{year}{2009}).

\bibitem[{\citenamefont{Fradelos and Wesolowski}(2011)}]{fradelos11}
\bibinfo{author}{\bibfnamefont{G.}~\bibnamefont{Fradelos}} \bibnamefont{and}
  \bibinfo{author}{\bibfnamefont{T.~A.} \bibnamefont{Wesolowski}},
  \bibinfo{journal}{J. Chem. Theory Comput.} \textbf{\bibinfo{volume}{7}},
  \bibinfo{pages}{213} (\bibinfo{year}{2011}).

\bibitem[{\citenamefont{Constantin
  et~al.}(2011{\natexlab{a}})\citenamefont{Constantin, Fabiano, Laricchia, and
  {Della Sala}}}]{apbe}
\bibinfo{author}{\bibfnamefont{L.~A.} \bibnamefont{Constantin}},
  \bibinfo{author}{\bibfnamefont{E.}~\bibnamefont{Fabiano}},
  \bibinfo{author}{\bibfnamefont{S.}~\bibnamefont{Laricchia}},
  \bibnamefont{and} \bibinfo{author}{\bibfnamefont{F.}~\bibnamefont{{Della
  Sala}}}, \bibinfo{journal}{Phys. Rev. Lett.} \textbf{\bibinfo{volume}{106}},
  \bibinfo{pages}{186406} (\bibinfo{year}{2011}{\natexlab{a}}).

\bibitem[{\citenamefont{Laricchia
  et~al.}(2011{\natexlab{a}})\citenamefont{Laricchia, Fabiano, Constantin, and
  {Della Sala}}}]{apbek}
\bibinfo{author}{\bibfnamefont{S.}~\bibnamefont{Laricchia}},
  \bibinfo{author}{\bibfnamefont{E.}~\bibnamefont{Fabiano}},
  \bibinfo{author}{\bibfnamefont{L.~A.} \bibnamefont{Constantin}},
  \bibnamefont{and} \bibinfo{author}{\bibfnamefont{F.}~\bibnamefont{{Della
  Sala}}}, \bibinfo{journal}{J. Chem. Theory Comput.}
  \textbf{\bibinfo{volume}{7}}, \bibinfo{pages}{2439}
  (\bibinfo{year}{2011}{\natexlab{a}}).

\bibitem[{\citenamefont{Laricchia
  et~al.}(2011{\natexlab{b}})\citenamefont{Laricchia, Fabiano, and {Della
  Sala}}}]{fde_lhf}
\bibinfo{author}{\bibfnamefont{S.}~\bibnamefont{Laricchia}},
  \bibinfo{author}{\bibfnamefont{E.}~\bibnamefont{Fabiano}}, \bibnamefont{and}
  \bibinfo{author}{\bibfnamefont{F.}~\bibnamefont{{Della Sala}}},
  \bibinfo{journal}{Chem. Phys. Lett.} \textbf{\bibinfo{volume}{518}},
  \bibinfo{pages}{114 } (\bibinfo{year}{2011}{\natexlab{b}}).

\bibitem[{\citenamefont{Laricchia et~al.}(2012)\citenamefont{Laricchia,
  Fabiano, and {Della Sala}}}]{fde_hyb_ene}
\bibinfo{author}{\bibfnamefont{S.}~\bibnamefont{Laricchia}},
  \bibinfo{author}{\bibfnamefont{E.}~\bibnamefont{Fabiano}}, \bibnamefont{and}
  \bibinfo{author}{\bibfnamefont{F.}~\bibnamefont{{Della Sala}}},
  \bibinfo{journal}{J. Chem. Phys.} \textbf{\bibinfo{volume}{137}},
  \bibinfo{eid}{014102} (\bibinfo{year}{2012}).

\bibitem[{\citenamefont{Fabiano
  et~al.}(2014{\natexlab{a}})\citenamefont{Fabiano, Laricchia, and {Della
  Sala}}}]{fdefract}
\bibinfo{author}{\bibfnamefont{E.}~\bibnamefont{Fabiano}},
  \bibinfo{author}{\bibfnamefont{S.}~\bibnamefont{Laricchia}},
  \bibnamefont{and} \bibinfo{author}{\bibfnamefont{F.}~\bibnamefont{{Della
  Sala}}}, \bibinfo{journal}{J. Chem. Phys.} \textbf{\bibinfo{volume}{140}},
  \bibinfo{eid}{114101} (\bibinfo{year}{2014}{\natexlab{a}}).

\bibitem[{\citenamefont{Laricchia et~al.}(2013)\citenamefont{Laricchia,
  Fabiano, and {Della Sala}}}]{fde_ct}
\bibinfo{author}{\bibfnamefont{S.}~\bibnamefont{Laricchia}},
  \bibinfo{author}{\bibfnamefont{E.}~\bibnamefont{Fabiano}}, \bibnamefont{and}
  \bibinfo{author}{\bibfnamefont{F.}~\bibnamefont{{Della Sala}}},
  \bibinfo{journal}{J. Chem. Phys.} \textbf{\bibinfo{volume}{138}},
  \bibinfo{eid}{124112} (\bibinfo{year}{2013}).

\bibitem[{\citenamefont{Laricchia et~al.}(2014)\citenamefont{Laricchia,
  Constantin, Fabiano, and {Della Sala}}}]{fde_lap}
\bibinfo{author}{\bibfnamefont{S.}~\bibnamefont{Laricchia}},
  \bibinfo{author}{\bibfnamefont{L.~A.} \bibnamefont{Constantin}},
  \bibinfo{author}{\bibfnamefont{E.}~\bibnamefont{Fabiano}}, \bibnamefont{and}
  \bibinfo{author}{\bibfnamefont{F.}~\bibnamefont{{Della Sala}}},
  \bibinfo{journal}{J. Chem. Theory Comput.} \textbf{\bibinfo{volume}{10}},
  \bibinfo{pages}{164} (\bibinfo{year}{2014}).

\bibitem[{\citenamefont{Schluns et~al.}(2015)\citenamefont{Schluns, Klahr,
  Muck-Lichtenfeld, Visscher, and Neugebauer}}]{schluns05}
\bibinfo{author}{\bibfnamefont{D.}~\bibnamefont{Schluns}},
  \bibinfo{author}{\bibfnamefont{K.}~\bibnamefont{Klahr}},
  \bibinfo{author}{\bibfnamefont{C.}~\bibnamefont{Muck-Lichtenfeld}},
  \bibinfo{author}{\bibfnamefont{L.}~\bibnamefont{Visscher}}, \bibnamefont{and}
  \bibinfo{author}{\bibfnamefont{J.}~\bibnamefont{Neugebauer}},
  \bibinfo{journal}{Phys. Chem. Chem. Phys.}  (\bibinfo{year}{2015}),
  \bibinfo{note}{in press}.

\bibitem[{\citenamefont{Kevorkyants et~al.}(2014)\citenamefont{Kevorkyants,
  Eshuis, and Pavanello}}]{pavanellovdw}
\bibinfo{author}{\bibfnamefont{R.}~\bibnamefont{Kevorkyants}},
  \bibinfo{author}{\bibfnamefont{H.}~\bibnamefont{Eshuis}}, \bibnamefont{and}
  \bibinfo{author}{\bibfnamefont{M.}~\bibnamefont{Pavanello}},
  \bibinfo{journal}{J. Chem. Phys.} \textbf{\bibinfo{volume}{141}},
  \bibinfo{eid}{044127} (\bibinfo{year}{2014}).

\bibitem[{\citenamefont{Jacob and Visscher}(2008)}]{jacob08prot}
\bibinfo{author}{\bibfnamefont{C.~R.} \bibnamefont{Jacob}} \bibnamefont{and}
  \bibinfo{author}{\bibfnamefont{L.}~\bibnamefont{Visscher}},
  \bibinfo{journal}{J. Chem. Phys.} \textbf{\bibinfo{volume}{128}},
  \bibinfo{pages}{155102} (\bibinfo{year}{2008}).

\bibitem[{\citenamefont{Fux et~al.}(2008)\citenamefont{Fux, Kiewisch, Jacob,
  Neugebauer, and Reiher}}]{fux08}
\bibinfo{author}{\bibfnamefont{S.}~\bibnamefont{Fux}},
  \bibinfo{author}{\bibfnamefont{K.}~\bibnamefont{Kiewisch}},
  \bibinfo{author}{\bibfnamefont{C.~R.} \bibnamefont{Jacob}},
  \bibinfo{author}{\bibfnamefont{J.}~\bibnamefont{Neugebauer}},
  \bibnamefont{and} \bibinfo{author}{\bibfnamefont{M.}~\bibnamefont{Reiher}},
  \bibinfo{journal}{Chem. Phys. Lett.} \textbf{\bibinfo{volume}{461}},
  \bibinfo{pages}{353 } (\bibinfo{year}{2008}).

\bibitem[{\citenamefont{Beyhan et~al.}(2010)\citenamefont{Beyhan, G\"{o}tz,
  Jacob, and Visscher}}]{beyhan10}
\bibinfo{author}{\bibfnamefont{S.~M.} \bibnamefont{Beyhan}},
  \bibinfo{author}{\bibfnamefont{A.~W.} \bibnamefont{G\"{o}tz}},
  \bibinfo{author}{\bibfnamefont{C.~R.} \bibnamefont{Jacob}}, \bibnamefont{and}
  \bibinfo{author}{\bibfnamefont{L.}~\bibnamefont{Visscher}},
  \bibinfo{journal}{J. Chem. Phys.} \textbf{\bibinfo{volume}{132}},
  \bibinfo{eid}{044114} (\bibinfo{year}{2010}).

\bibitem[{\citenamefont{Fux et~al.}(2010)\citenamefont{Fux, Jacob, Neugebauer,
  Visscher, and Reiher}}]{fux10}
\bibinfo{author}{\bibfnamefont{S.}~\bibnamefont{Fux}},
  \bibinfo{author}{\bibfnamefont{C.~R.} \bibnamefont{Jacob}},
  \bibinfo{author}{\bibfnamefont{J.}~\bibnamefont{Neugebauer}},
  \bibinfo{author}{\bibfnamefont{L.}~\bibnamefont{Visscher}}, \bibnamefont{and}
  \bibinfo{author}{\bibfnamefont{M.}~\bibnamefont{Reiher}},
  \bibinfo{journal}{J. Chem. Phys.} \textbf{\bibinfo{volume}{132}},
  \bibinfo{eid}{164101} (\bibinfo{year}{2010}).

\bibitem[{\citenamefont{Wesolowski and Warshel}(1993)}]{wesowarh93}
\bibinfo{author}{\bibfnamefont{T.~A.} \bibnamefont{Wesolowski}}
  \bibnamefont{and} \bibinfo{author}{\bibfnamefont{A.}~\bibnamefont{Warshel}},
  \bibinfo{journal}{J. Phys. Chem.} \textbf{\bibinfo{volume}{97}},
  \bibinfo{pages}{8050} (\bibinfo{year}{1993}).

\bibitem[{\citenamefont{Hodak et~al.}(2008)\citenamefont{Hodak, Lu, and
  Bernholc}}]{hodak08}
\bibinfo{author}{\bibfnamefont{M.}~\bibnamefont{Hodak}},
  \bibinfo{author}{\bibfnamefont{W.}~\bibnamefont{Lu}}, \bibnamefont{and}
  \bibinfo{author}{\bibfnamefont{J.}~\bibnamefont{Bernholc}},
  \bibinfo{journal}{J. Chem. Phys.} \textbf{\bibinfo{volume}{128}},
  \bibinfo{eid}{014101} (\bibinfo{year}{2008}).

\bibitem[{\citenamefont{Neugebauer
  et~al.}(2005{\natexlab{a}})\citenamefont{Neugebauer, Louwerse, Baerends, and
  Wesolowski}}]{neugebauer05solv}
\bibinfo{author}{\bibfnamefont{J.}~\bibnamefont{Neugebauer}},
  \bibinfo{author}{\bibfnamefont{M.~J.} \bibnamefont{Louwerse}},
  \bibinfo{author}{\bibfnamefont{E.~J.} \bibnamefont{Baerends}},
  \bibnamefont{and} \bibinfo{author}{\bibfnamefont{T.~A.}
  \bibnamefont{Wesolowski}}, \bibinfo{journal}{J. Chem. Phys.}
  \textbf{\bibinfo{volume}{122}}, \bibinfo{eid}{094115}
  (\bibinfo{year}{2005}{\natexlab{a}}).

\bibitem[{\citenamefont{Jacob et~al.}(2006)\citenamefont{Jacob, Neugebauer,
  Jensen, and Visscher}}]{jacob06solv}
\bibinfo{author}{\bibfnamefont{C.~R.} \bibnamefont{Jacob}},
  \bibinfo{author}{\bibfnamefont{J.}~\bibnamefont{Neugebauer}},
  \bibinfo{author}{\bibfnamefont{L.}~\bibnamefont{Jensen}}, \bibnamefont{and}
  \bibinfo{author}{\bibfnamefont{L.}~\bibnamefont{Visscher}},
  \bibinfo{journal}{Phys. Chem. Chem. Phys.} \textbf{\bibinfo{volume}{8}},
  \bibinfo{pages}{2349} (\bibinfo{year}{2006}).

\bibitem[{\citenamefont{Neugebauer
  et~al.}(2005{\natexlab{b}})\citenamefont{Neugebauer, Louwerse, Belanzoni,
  Wesolowski, and Baerends}}]{neugebauer05spin}
\bibinfo{author}{\bibfnamefont{J.}~\bibnamefont{Neugebauer}},
  \bibinfo{author}{\bibfnamefont{M.~J.} \bibnamefont{Louwerse}},
  \bibinfo{author}{\bibfnamefont{P.}~\bibnamefont{Belanzoni}},
  \bibinfo{author}{\bibfnamefont{T.~A.} \bibnamefont{Wesolowski}},
  \bibnamefont{and} \bibinfo{author}{\bibfnamefont{E.~J.}
  \bibnamefont{Baerends}}, \bibinfo{journal}{J. Chem. Phys.}
  \textbf{\bibinfo{volume}{123}}, \bibinfo{eid}{114101}
  (\bibinfo{year}{2005}{\natexlab{b}}).

\bibitem[{\citenamefont{Kaminski et~al.}(2010)\citenamefont{Kaminski, Gusarov,
  Wesolowski, and Kovalenko}}]{kaminski10}
\bibinfo{author}{\bibfnamefont{J.~W.} \bibnamefont{Kaminski}},
  \bibinfo{author}{\bibfnamefont{S.}~\bibnamefont{Gusarov}},
  \bibinfo{author}{\bibfnamefont{T.~A.} \bibnamefont{Wesolowski}},
  \bibnamefont{and}
  \bibinfo{author}{\bibfnamefont{A.}~\bibnamefont{Kovalenko}},
  \bibinfo{journal}{J. Phys. Chem. A} \textbf{\bibinfo{volume}{114}},
  \bibinfo{pages}{6082} (\bibinfo{year}{2010}).

\bibitem[{\citenamefont{Kiewisch et~al.}(2013)\citenamefont{Kiewisch, Jacob,
  and Visscher}}]{kiewisch13}
\bibinfo{author}{\bibfnamefont{K.}~\bibnamefont{Kiewisch}},
  \bibinfo{author}{\bibfnamefont{C.~R.} \bibnamefont{Jacob}}, \bibnamefont{and}
  \bibinfo{author}{\bibfnamefont{L.}~\bibnamefont{Visscher}},
  \bibinfo{journal}{J. Chem. Theory Comput.} \textbf{\bibinfo{volume}{9}},
  \bibinfo{pages}{2425} (\bibinfo{year}{2013}).

\bibitem[{\citenamefont{Tran et~al.}(2002)\citenamefont{Tran, Weber,
  Wesołowski, Cheikh, Ellinger, and Pauzat}}]{tran02}
\bibinfo{author}{\bibfnamefont{F.}~\bibnamefont{Tran}},
  \bibinfo{author}{\bibfnamefont{J.}~\bibnamefont{Weber}},
  \bibinfo{author}{\bibfnamefont{T.~A.} \bibnamefont{Wesołowski}},
  \bibinfo{author}{\bibfnamefont{F.}~\bibnamefont{Cheikh}},
  \bibinfo{author}{\bibfnamefont{Y.}~\bibnamefont{Ellinger}}, \bibnamefont{and}
  \bibinfo{author}{\bibfnamefont{F.}~\bibnamefont{Pauzat}},
  \bibinfo{journal}{J. Phys. Chem. B} \textbf{\bibinfo{volume}{106}},
  \bibinfo{pages}{8689} (\bibinfo{year}{2002}).

\bibitem[{\citenamefont{Lembarki and Chermette}(1994)}]{lc94}
\bibinfo{author}{\bibfnamefont{A.}~\bibnamefont{Lembarki}} \bibnamefont{and}
  \bibinfo{author}{\bibfnamefont{H.}~\bibnamefont{Chermette}},
  \bibinfo{journal}{Phys. Rev. A} \textbf{\bibinfo{volume}{50}},
  \bibinfo{pages}{5328} (\bibinfo{year}{1994}).

\bibitem[{\citenamefont{Thakkar}(1992)}]{thak92}
\bibinfo{author}{\bibfnamefont{A.~J.} \bibnamefont{Thakkar}},
  \bibinfo{journal}{Phys. Rev. A} \textbf{\bibinfo{volume}{46}},
  \bibinfo{pages}{6920} (\bibinfo{year}{1992}).

\bibitem[{\citenamefont{Lee et~al.}(1991)\citenamefont{Lee, Lee, and
  Parr}}]{llp91}
\bibinfo{author}{\bibfnamefont{H.}~\bibnamefont{Lee}},
  \bibinfo{author}{\bibfnamefont{C.}~\bibnamefont{Lee}}, \bibnamefont{and}
  \bibinfo{author}{\bibfnamefont{R.~G.} \bibnamefont{Parr}},
  \bibinfo{journal}{Phys. Rev. A} \textbf{\bibinfo{volume}{44}},
  \bibinfo{pages}{768} (\bibinfo{year}{1991}).

\bibitem[{\citenamefont{Tran and Wesolowski}(2013)}]{weso_chap}
\bibinfo{author}{\bibfnamefont{F.}~\bibnamefont{Tran}} \bibnamefont{and}
  \bibinfo{author}{\bibfnamefont{T.~A.} \bibnamefont{Wesolowski}}, in
  \emph{\bibinfo{booktitle}{Recent Progress in Orbital-free Density Functional
  Theory}}, edited by \bibinfo{editor}{\bibfnamefont{T.~A.}
  \bibnamefont{Wesolowsky}} \bibnamefont{and}
  \bibinfo{editor}{\bibfnamefont{Y.~A.} \bibnamefont{Wang}}
  (\bibinfo{publisher}{World Scientific}, \bibinfo{address}{Singapore},
  \bibinfo{year}{2013}), pp. \bibinfo{pages}{429--442}.

\bibitem[{\citenamefont{Laricchia et~al.}(2010)\citenamefont{Laricchia,
  Fabiano, and {Della Sala}}}]{fde_hybrid}
\bibinfo{author}{\bibfnamefont{S.}~\bibnamefont{Laricchia}},
  \bibinfo{author}{\bibfnamefont{E.}~\bibnamefont{Fabiano}}, \bibnamefont{and}
  \bibinfo{author}{\bibfnamefont{F.}~\bibnamefont{{Della Sala}}},
  \bibinfo{journal}{J. Chem. Phys.} \textbf{\bibinfo{volume}{133}},
  \bibinfo{eid}{164111} (\bibinfo{year}{2010}).

\bibitem[{\citenamefont{Weso\l{}owski}(2008)}]{weso08det}
\bibinfo{author}{\bibfnamefont{T.~A.} \bibnamefont{Weso\l{}owski}},
  \bibinfo{journal}{Phys. Rev. A} \textbf{\bibinfo{volume}{77}},
  \bibinfo{pages}{012504} (\bibinfo{year}{2008}).

\bibitem[{\citenamefont{Jacob et~al.}(2005)\citenamefont{Jacob, Wesolowski, and
  Visscher}}]{jacob05dip}
\bibinfo{author}{\bibfnamefont{C.~R.} \bibnamefont{Jacob}},
  \bibinfo{author}{\bibfnamefont{T.~A.} \bibnamefont{Wesolowski}},
  \bibnamefont{and} \bibinfo{author}{\bibfnamefont{L.}~\bibnamefont{Visscher}},
  \bibinfo{journal}{J. Chem. Phys.} \textbf{\bibinfo{volume}{123}},
  \bibinfo{eid}{174104} (\bibinfo{year}{2005}).

\bibitem[{\citenamefont{Pernal and Wesolowski}(2009)}]{pernalweso09}
\bibinfo{author}{\bibfnamefont{K.}~\bibnamefont{Pernal}} \bibnamefont{and}
  \bibinfo{author}{\bibfnamefont{T.}~\bibnamefont{Wesolowski}},
  \bibinfo{journal}{Int. Jou. Quant. Chem.} \textbf{\bibinfo{volume}{109}},
  \bibinfo{pages}{2520} (\bibinfo{year}{2009}).

\bibitem[{\citenamefont{Tao et~al.}(2003)\citenamefont{Tao, Perdew, Staroverov,
  and Scuseria}}]{tpss}
\bibinfo{author}{\bibfnamefont{J.}~\bibnamefont{Tao}},
  \bibinfo{author}{\bibfnamefont{J.~P.} \bibnamefont{Perdew}},
  \bibinfo{author}{\bibfnamefont{V.~N.} \bibnamefont{Staroverov}},
  \bibnamefont{and} \bibinfo{author}{\bibfnamefont{G.~E.}
  \bibnamefont{Scuseria}}, \bibinfo{journal}{Phys. Rev. Lett.}
  \textbf{\bibinfo{volume}{91}}, \bibinfo{pages}{146401}
  (\bibinfo{year}{2003}).

\bibitem[{\citenamefont{Perdew et~al.}(2009)\citenamefont{Perdew, Ruzsinszky,
  Csonka, Constantin, and Sun}}]{revtpss}
\bibinfo{author}{\bibfnamefont{J.~P.} \bibnamefont{Perdew}},
  \bibinfo{author}{\bibfnamefont{A.}~\bibnamefont{Ruzsinszky}},
  \bibinfo{author}{\bibfnamefont{G.~I.} \bibnamefont{Csonka}},
  \bibinfo{author}{\bibfnamefont{L.~A.} \bibnamefont{Constantin}},
  \bibnamefont{and} \bibinfo{author}{\bibfnamefont{J.}~\bibnamefont{Sun}},
  \bibinfo{journal}{Phys. Rev. Lett.} \textbf{\bibinfo{volume}{103}},
  \bibinfo{pages}{026403} (\bibinfo{year}{2009}).

\bibitem[{\citenamefont{Constantin
  et~al.}(2011{\natexlab{b}})\citenamefont{Constantin, Chiodo, Fabiano,
  Bodrenko, and {Della Sala}}}]{js}
\bibinfo{author}{\bibfnamefont{L.~A.} \bibnamefont{Constantin}},
  \bibinfo{author}{\bibfnamefont{L.}~\bibnamefont{Chiodo}},
  \bibinfo{author}{\bibfnamefont{E.}~\bibnamefont{Fabiano}},
  \bibinfo{author}{\bibfnamefont{I.}~\bibnamefont{Bodrenko}}, \bibnamefont{and}
  \bibinfo{author}{\bibfnamefont{F.}~\bibnamefont{{Della Sala}}},
  \bibinfo{journal}{Phys. Rev. B} \textbf{\bibinfo{volume}{84}},
  \bibinfo{pages}{045126} (\bibinfo{year}{2011}{\natexlab{b}}).

\bibitem[{\citenamefont{Constantin et~al.}(2012)\citenamefont{Constantin,
  Fabiano, and {Della Sala}}}]{tpssloc}
\bibinfo{author}{\bibfnamefont{L.~A.} \bibnamefont{Constantin}},
  \bibinfo{author}{\bibfnamefont{E.}~\bibnamefont{Fabiano}}, \bibnamefont{and}
  \bibinfo{author}{\bibfnamefont{F.}~\bibnamefont{{Della Sala}}},
  \bibinfo{journal}{Phys. Rev. B} \textbf{\bibinfo{volume}{86}},
  \bibinfo{pages}{035130} (\bibinfo{year}{2012}).

\bibitem[{\citenamefont{Constantin
  et~al.}(2013{\natexlab{a}})\citenamefont{Constantin, Fabiano, and {Della
  Sala}}}]{bloc}
\bibinfo{author}{\bibfnamefont{L.~A.} \bibnamefont{Constantin}},
  \bibinfo{author}{\bibfnamefont{E.}~\bibnamefont{Fabiano}}, \bibnamefont{and}
  \bibinfo{author}{\bibfnamefont{F.}~\bibnamefont{{Della Sala}}},
  \bibinfo{journal}{J. Chem. Theory Comput.} \textbf{\bibinfo{volume}{9}},
  \bibinfo{pages}{2256} (\bibinfo{year}{2013}{\natexlab{a}}).

\bibitem[{\citenamefont{Van~Voorhis and Scuseria}(1998)}]{vsxc}
\bibinfo{author}{\bibfnamefont{T.}~\bibnamefont{Van~Voorhis}} \bibnamefont{and}
  \bibinfo{author}{\bibfnamefont{G.~E.} \bibnamefont{Scuseria}},
  \bibinfo{journal}{J. Chem. Phys.} \textbf{\bibinfo{volume}{109}},
  \bibinfo{pages}{400} (\bibinfo{year}{1998}).

\bibitem[{\citenamefont{Schmider and Becke}(1998)}]{477481}
\bibinfo{author}{\bibfnamefont{H.~L.} \bibnamefont{Schmider}} \bibnamefont{and}
  \bibinfo{author}{\bibfnamefont{A.~D.} \bibnamefont{Becke}},
  \bibinfo{journal}{J. Chem. Phys.} \textbf{\bibinfo{volume}{109}},
  \bibinfo{pages}{8188} (\bibinfo{year}{1998}).

\bibitem[{\citenamefont{Zhao and Truhlar}(2006)}]{m06l}
\bibinfo{author}{\bibfnamefont{Y.}~\bibnamefont{Zhao}} \bibnamefont{and}
  \bibinfo{author}{\bibfnamefont{D.~G.} \bibnamefont{Truhlar}},
  \bibinfo{journal}{J. Chem. Phys.} \textbf{\bibinfo{volume}{125}},
  \bibinfo{eid}{194101} (\bibinfo{year}{2006}).

\bibitem[{\citenamefont{Peverati and Truhlar}(2012)}]{m11l}
\bibinfo{author}{\bibfnamefont{R.}~\bibnamefont{Peverati}} \bibnamefont{and}
  \bibinfo{author}{\bibfnamefont{D.~G.} \bibnamefont{Truhlar}},
  \bibinfo{journal}{J. Phys. Chem. Lett.} \textbf{\bibinfo{volume}{3}},
  \bibinfo{pages}{117} (\bibinfo{year}{2012}).

\bibitem[{\citenamefont{Ruzsinszky et~al.}(2012)\citenamefont{Ruzsinszky, Sun,
  Xiao, and Csonka}}]{regtpss}
\bibinfo{author}{\bibfnamefont{A.}~\bibnamefont{Ruzsinszky}},
  \bibinfo{author}{\bibfnamefont{J.}~\bibnamefont{Sun}},
  \bibinfo{author}{\bibfnamefont{B.}~\bibnamefont{Xiao}}, \bibnamefont{and}
  \bibinfo{author}{\bibfnamefont{G.~I.} \bibnamefont{Csonka}},
  \bibinfo{journal}{J. Chem. Theory Comput.} \textbf{\bibinfo{volume}{8}},
  \bibinfo{pages}{2078} (\bibinfo{year}{2012}).

\bibitem[{\citenamefont{Sun et~al.}(2012)\citenamefont{Sun, Xiao, and
  Ruzsinszky}}]{mggms_1}
\bibinfo{author}{\bibfnamefont{J.}~\bibnamefont{Sun}},
  \bibinfo{author}{\bibfnamefont{B.}~\bibnamefont{Xiao}}, \bibnamefont{and}
  \bibinfo{author}{\bibfnamefont{A.}~\bibnamefont{Ruzsinszky}},
  \bibinfo{journal}{J. Chem. Phys.} \textbf{\bibinfo{volume}{137}},
  \bibinfo{eid}{051101} (\bibinfo{year}{2012}).

\bibitem[{\citenamefont{Sun et~al.}(2013{\natexlab{a}})\citenamefont{Sun,
  Haunschild, Xiao, Bulik, Scuseria, and Perdew}}]{mgga_ms2}
\bibinfo{author}{\bibfnamefont{J.}~\bibnamefont{Sun}},
  \bibinfo{author}{\bibfnamefont{R.}~\bibnamefont{Haunschild}},
  \bibinfo{author}{\bibfnamefont{B.}~\bibnamefont{Xiao}},
  \bibinfo{author}{\bibfnamefont{I.~W.} \bibnamefont{Bulik}},
  \bibinfo{author}{\bibfnamefont{G.~E.} \bibnamefont{Scuseria}},
  \bibnamefont{and} \bibinfo{author}{\bibfnamefont{J.~P.}
  \bibnamefont{Perdew}}, \bibinfo{journal}{J. Chem. Phys.}
  \textbf{\bibinfo{volume}{138}}, \bibinfo{eid}{044113}
  (\bibinfo{year}{2013}{\natexlab{a}}).

\bibitem[{\citenamefont{Sun et~al.}(2015)\citenamefont{Sun, Perdew, and
  Ruzsinszky}}]{sunpnas}
\bibinfo{author}{\bibfnamefont{J.}~\bibnamefont{Sun}},
  \bibinfo{author}{\bibfnamefont{J.~P.} \bibnamefont{Perdew}},
  \bibnamefont{and}
  \bibinfo{author}{\bibfnamefont{A.}~\bibnamefont{Ruzsinszky}},
  \bibinfo{journal}{Proc. Nat. Acad. Sci.} \textbf{\bibinfo{volume}{112}},
  \bibinfo{pages}{685} (\bibinfo{year}{2015}).

\bibitem[{\citenamefont{Wellendorff et~al.}(2014)\citenamefont{Wellendorff,
  Lundgaard, Jacobsen, and Bligaard}}]{beefmeta14}
\bibinfo{author}{\bibfnamefont{J.}~\bibnamefont{Wellendorff}},
  \bibinfo{author}{\bibfnamefont{K.~T.} \bibnamefont{Lundgaard}},
  \bibinfo{author}{\bibfnamefont{K.~W.} \bibnamefont{Jacobsen}},
  \bibnamefont{and} \bibinfo{author}{\bibfnamefont{T.}~\bibnamefont{Bligaard}},
  \bibinfo{journal}{J. Chem. Phys.} \textbf{\bibinfo{volume}{140}},
  \bibinfo{eid}{144107} (\bibinfo{year}{2014}).

\bibitem[{\citenamefont{Mardirossian and Head-Gordon}(2015)}]{b97mv2015}
\bibinfo{author}{\bibfnamefont{N.}~\bibnamefont{Mardirossian}}
  \bibnamefont{and}
  \bibinfo{author}{\bibfnamefont{M.}~\bibnamefont{Head-Gordon}},
  \bibinfo{journal}{J. Chem. Phys.} \textbf{\bibinfo{volume}{142}},
  \bibinfo{eid}{074111} (\bibinfo{year}{2015}).

\bibitem[{\citenamefont{Constantin
  et~al.}(2013{\natexlab{b}})\citenamefont{Constantin, Fabiano, and {Della
  Sala}}}]{bloc_hole}
\bibinfo{author}{\bibfnamefont{L.~A.} \bibnamefont{Constantin}},
  \bibinfo{author}{\bibfnamefont{E.}~\bibnamefont{Fabiano}}, \bibnamefont{and}
  \bibinfo{author}{\bibfnamefont{F.}~\bibnamefont{{Della Sala}}},
  \bibinfo{journal}{Phys. Rev. B} \textbf{\bibinfo{volume}{88}},
  \bibinfo{pages}{125112} (\bibinfo{year}{2013}{\natexlab{b}}).

\bibitem[{\citenamefont{Della~Sala et~al.}(2015)\citenamefont{Della~Sala,
  Fabiano, and Constantin}}]{fkato}
\bibinfo{author}{\bibfnamefont{F.}~\bibnamefont{Della~Sala}},
  \bibinfo{author}{\bibfnamefont{E.}~\bibnamefont{Fabiano}}, \bibnamefont{and}
  \bibinfo{author}{\bibfnamefont{L.~A.} \bibnamefont{Constantin}},
  \bibinfo{journal}{Phys. Rev. B} \textbf{\bibinfo{volume}{91}},
  \bibinfo{pages}{035126} (\bibinfo{year}{2015}).

\bibitem[{\citenamefont{Xiao et~al.}(2013)\citenamefont{Xiao, Sun, Ruzsinszky,
  Feng, Haunschild, Scuseria, and Perdew}}]{xiao13}
\bibinfo{author}{\bibfnamefont{B.}~\bibnamefont{Xiao}},
  \bibinfo{author}{\bibfnamefont{J.}~\bibnamefont{Sun}},
  \bibinfo{author}{\bibfnamefont{A.}~\bibnamefont{Ruzsinszky}},
  \bibinfo{author}{\bibfnamefont{J.}~\bibnamefont{Feng}},
  \bibinfo{author}{\bibfnamefont{R.}~\bibnamefont{Haunschild}},
  \bibinfo{author}{\bibfnamefont{G.~E.} \bibnamefont{Scuseria}},
  \bibnamefont{and} \bibinfo{author}{\bibfnamefont{J.~P.}
  \bibnamefont{Perdew}}, \bibinfo{journal}{Phys. Rev. B}
  \textbf{\bibinfo{volume}{88}}, \bibinfo{pages}{184103}
  (\bibinfo{year}{2013}).

\bibitem[{\citenamefont{Sun et~al.}(2013{\natexlab{b}})\citenamefont{Sun, Xiao,
  Fang, Haunschild, Hao, Ruzsinszky, Csonka, Scuseria, and Perdew}}]{sunprl}
\bibinfo{author}{\bibfnamefont{J.}~\bibnamefont{Sun}},
  \bibinfo{author}{\bibfnamefont{B.}~\bibnamefont{Xiao}},
  \bibinfo{author}{\bibfnamefont{Y.}~\bibnamefont{Fang}},
  \bibinfo{author}{\bibfnamefont{R.}~\bibnamefont{Haunschild}},
  \bibinfo{author}{\bibfnamefont{P.}~\bibnamefont{Hao}},
  \bibinfo{author}{\bibfnamefont{A.}~\bibnamefont{Ruzsinszky}},
  \bibinfo{author}{\bibfnamefont{G.~I.} \bibnamefont{Csonka}},
  \bibinfo{author}{\bibfnamefont{G.~E.} \bibnamefont{Scuseria}},
  \bibnamefont{and} \bibinfo{author}{\bibfnamefont{J.~P.}
  \bibnamefont{Perdew}}, \bibinfo{journal}{Phys. Rev. Lett.}
  \textbf{\bibinfo{volume}{111}}, \bibinfo{pages}{106401}
  (\bibinfo{year}{2013}{\natexlab{b}}).

\bibitem[{\citenamefont{Staroverov et~al.}(2004)\citenamefont{Staroverov,
  Scuseria, Tao, and Perdew}}]{stare04}
\bibinfo{author}{\bibfnamefont{V.~N.} \bibnamefont{Staroverov}},
  \bibinfo{author}{\bibfnamefont{G.~E.} \bibnamefont{Scuseria}},
  \bibinfo{author}{\bibfnamefont{J.}~\bibnamefont{Tao}}, \bibnamefont{and}
  \bibinfo{author}{\bibfnamefont{J.~P.} \bibnamefont{Perdew}},
  \bibinfo{journal}{Phys. Rev. B} \textbf{\bibinfo{volume}{69}},
  \bibinfo{pages}{075102} (\bibinfo{year}{2004}).

\bibitem[{\citenamefont{Adamo et~al.}(2000)\citenamefont{Adamo, Ernzerhof, and
  Scuseria}}]{adam00}
\bibinfo{author}{\bibfnamefont{C.}~\bibnamefont{Adamo}},
  \bibinfo{author}{\bibfnamefont{M.}~\bibnamefont{Ernzerhof}},
  \bibnamefont{and} \bibinfo{author}{\bibfnamefont{G.~E.}
  \bibnamefont{Scuseria}}, \bibinfo{journal}{J. Chem. Phys.}
  \textbf{\bibinfo{volume}{112}}, \bibinfo{pages}{2643} (\bibinfo{year}{2000}).

\bibitem[{\citenamefont{Riley et~al.}(2007)\citenamefont{Riley, Op't~Holt, and
  Merz}}]{riley07}
\bibinfo{author}{\bibfnamefont{K.~E.} \bibnamefont{Riley}},
  \bibinfo{author}{\bibfnamefont{B.~T.} \bibnamefont{Op't~Holt}},
  \bibnamefont{and} \bibinfo{author}{\bibfnamefont{K.~M.} \bibnamefont{Merz}},
  \bibinfo{journal}{J. Chem. Theory Comput.} \textbf{\bibinfo{volume}{3}},
  \bibinfo{pages}{407} (\bibinfo{year}{2007}).

\bibitem[{\citenamefont{Andersen et~al.}(2012)\citenamefont{Andersen,
  Hornek\ae{}r, and Hammer}}]{hammer12}
\bibinfo{author}{\bibfnamefont{M.}~\bibnamefont{Andersen}},
  \bibinfo{author}{\bibfnamefont{L.}~\bibnamefont{Hornek\ae{}r}},
  \bibnamefont{and} \bibinfo{author}{\bibfnamefont{B.}~\bibnamefont{Hammer}},
  \bibinfo{journal}{Phys. Rev. B} \textbf{\bibinfo{volume}{86}},
  \bibinfo{pages}{085405} (\bibinfo{year}{2012}).

\bibitem[{\citenamefont{Luo et~al.}(2012)\citenamefont{Luo, Zhao, and
  Truhlar}}]{m06surf}
\bibinfo{author}{\bibfnamefont{S.}~\bibnamefont{Luo}},
  \bibinfo{author}{\bibfnamefont{Y.}~\bibnamefont{Zhao}}, \bibnamefont{and}
  \bibinfo{author}{\bibfnamefont{D.~G.} \bibnamefont{Truhlar}},
  \bibinfo{journal}{J. Phys. Chem. Lett.} \textbf{\bibinfo{volume}{3}},
  \bibinfo{pages}{2975} (\bibinfo{year}{2012}).

\bibitem[{\citenamefont{Sun et~al.}(2011)\citenamefont{Sun, Marsman,
  Ruzsinszky, Kresse, and Perdew}}]{sunmeta11}
\bibinfo{author}{\bibfnamefont{J.}~\bibnamefont{Sun}},
  \bibinfo{author}{\bibfnamefont{M.}~\bibnamefont{Marsman}},
  \bibinfo{author}{\bibfnamefont{A.}~\bibnamefont{Ruzsinszky}},
  \bibinfo{author}{\bibfnamefont{G.}~\bibnamefont{Kresse}}, \bibnamefont{and}
  \bibinfo{author}{\bibfnamefont{J.~P.} \bibnamefont{Perdew}},
  \bibinfo{journal}{Phys. Rev. B} \textbf{\bibinfo{volume}{83}},
  \bibinfo{pages}{121410} (\bibinfo{year}{2011}).

\bibitem[{\citenamefont{Hao et~al.}(2013)\citenamefont{Hao, Sun, Xiao,
  Ruzsinszky, Csonka, Tao, Glindmeyer, and Perdew}}]{hao13}
\bibinfo{author}{\bibfnamefont{P.}~\bibnamefont{Hao}},
  \bibinfo{author}{\bibfnamefont{J.}~\bibnamefont{Sun}},
  \bibinfo{author}{\bibfnamefont{B.}~\bibnamefont{Xiao}},
  \bibinfo{author}{\bibfnamefont{A.}~\bibnamefont{Ruzsinszky}},
  \bibinfo{author}{\bibfnamefont{G.~I.} \bibnamefont{Csonka}},
  \bibinfo{author}{\bibfnamefont{J.}~\bibnamefont{Tao}},
  \bibinfo{author}{\bibfnamefont{S.}~\bibnamefont{Glindmeyer}},
  \bibnamefont{and} \bibinfo{author}{\bibfnamefont{J.~P.}
  \bibnamefont{Perdew}}, \bibinfo{journal}{J. Chem. Theory Comput.}
  \textbf{\bibinfo{volume}{9}}, \bibinfo{pages}{355} (\bibinfo{year}{2013}).

\bibitem[{\citenamefont{Fabiano
  et~al.}(2014{\natexlab{b}})\citenamefont{Fabiano, Constantin, and
  Della~Sala}}]{dihyd14}
\bibinfo{author}{\bibfnamefont{E.}~\bibnamefont{Fabiano}},
  \bibinfo{author}{\bibfnamefont{L.~A.} \bibnamefont{Constantin}},
  \bibnamefont{and}
  \bibinfo{author}{\bibfnamefont{F.}~\bibnamefont{Della~Sala}},
  \bibinfo{journal}{J. Chem. Theory Comput.} \textbf{\bibinfo{volume}{10}},
  \bibinfo{pages}{3151} (\bibinfo{year}{2014}{\natexlab{b}}).

\bibitem[{\citenamefont{Nazarov and Vignale}(2011)}]{vignale}
\bibinfo{author}{\bibfnamefont{V.~U.} \bibnamefont{Nazarov}} \bibnamefont{and}
  \bibinfo{author}{\bibfnamefont{G.}~\bibnamefont{Vignale}},
  \bibinfo{journal}{Phys. Rev. Lett.} \textbf{\bibinfo{volume}{107}},
  \bibinfo{pages}{216402} (\bibinfo{year}{2011}).

\bibitem[{\citenamefont{Arbuznikov and
  Kaupp}(2003{\natexlab{a}})}]{arbuznikov03}
\bibinfo{author}{\bibfnamefont{A.~V.} \bibnamefont{Arbuznikov}}
  \bibnamefont{and} \bibinfo{author}{\bibfnamefont{M.}~\bibnamefont{Kaupp}},
  \bibinfo{journal}{Chem. Phys. Lett.} \textbf{\bibinfo{volume}{381}},
  \bibinfo{pages}{495 } (\bibinfo{year}{2003}{\natexlab{a}}).

\bibitem[{\citenamefont{Seidl et~al.}(1996)\citenamefont{Seidl, G\"orling,
  Vogl, Majewski, and Levy}}]{seidl96}
\bibinfo{author}{\bibfnamefont{A.}~\bibnamefont{Seidl}},
  \bibinfo{author}{\bibfnamefont{A.}~\bibnamefont{G\"orling}},
  \bibinfo{author}{\bibfnamefont{P.}~\bibnamefont{Vogl}},
  \bibinfo{author}{\bibfnamefont{J.~A.} \bibnamefont{Majewski}},
  \bibnamefont{and} \bibinfo{author}{\bibfnamefont{M.}~\bibnamefont{Levy}},
  \bibinfo{journal}{Phys. Rev. B} \textbf{\bibinfo{volume}{53}},
  \bibinfo{pages}{3764} (\bibinfo{year}{1996}).

\bibitem[{\citenamefont{Gritsenko}(2013)}]{gritchap}
\bibinfo{author}{\bibfnamefont{O.}~\bibnamefont{Gritsenko}}, in
  \emph{\bibinfo{booktitle}{Recent Progress in Orbital-free Density Functional
  Theory}}, edited by \bibinfo{editor}{\bibfnamefont{T.~A.}
  \bibnamefont{Wesolowsky}} \bibnamefont{and}
  \bibinfo{editor}{\bibfnamefont{Y.~A.} \bibnamefont{Wang}}
  (\bibinfo{publisher}{World Scientific}, \bibinfo{address}{Singapore},
  \bibinfo{year}{2013}), pp. \bibinfo{pages}{355--365}.

\bibitem[{\citenamefont{Arbuznikov and
  Kaupp}(2003{\natexlab{b}})}]{Arbuznikov2003495}
\bibinfo{author}{\bibfnamefont{A.~V.} \bibnamefont{Arbuznikov}}
  \bibnamefont{and} \bibinfo{author}{\bibfnamefont{M.}~\bibnamefont{Kaupp}},
  \bibinfo{journal}{Chem. Phys. Lett.} \textbf{\bibinfo{volume}{381}},
  \bibinfo{pages}{495 } (\bibinfo{year}{2003}{\natexlab{b}}).

\bibitem[{\citenamefont{Arbuznikov et~al.}(2002)\citenamefont{Arbuznikov,
  Kaupp, Malkin, Reviakine, and Malkina}}]{B207171A}
\bibinfo{author}{\bibfnamefont{A.~V.} \bibnamefont{Arbuznikov}},
  \bibinfo{author}{\bibfnamefont{M.}~\bibnamefont{Kaupp}},
  \bibinfo{author}{\bibfnamefont{V.~G.} \bibnamefont{Malkin}},
  \bibinfo{author}{\bibfnamefont{R.}~\bibnamefont{Reviakine}},
  \bibnamefont{and} \bibinfo{author}{\bibfnamefont{O.~L.}
  \bibnamefont{Malkina}}, \bibinfo{journal}{Phys. Chem. Chem. Phys.}
  \textbf{\bibinfo{volume}{4}}, \bibinfo{pages}{5467} (\bibinfo{year}{2002}).

\bibitem[{\citenamefont{Humbert-Droz et~al.}(2013)\citenamefont{Humbert-Droz,
  Zhou, Shedge, and Wesolowski}}]{humbert13}
\bibinfo{author}{\bibfnamefont{M.}~\bibnamefont{Humbert-Droz}},
  \bibinfo{author}{\bibfnamefont{X.}~\bibnamefont{Zhou}},
  \bibinfo{author}{\bibfnamefont{S.}~\bibnamefont{Shedge}}, \bibnamefont{and}
  \bibinfo{author}{\bibfnamefont{T.}~\bibnamefont{Wesolowski}},
  \bibinfo{journal}{Theor. Chem. Acc.} \textbf{\bibinfo{volume}{133}},
  \bibinfo{eid}{1405} (\bibinfo{year}{2013}).

\bibitem[{\citenamefont{Nafziger and Wasserman}(2014)}]{wassermanrev}
\bibinfo{author}{\bibfnamefont{J.}~\bibnamefont{Nafziger}} \bibnamefont{and}
  \bibinfo{author}{\bibfnamefont{A.}~\bibnamefont{Wasserman}},
  \bibinfo{journal}{J. Phys. Chem. A} \textbf{\bibinfo{volume}{118}},
  \bibinfo{pages}{7623} (\bibinfo{year}{2014}).

\bibitem[{\citenamefont{Zhao et~al.}(1995)\citenamefont{Zhao, Morrison, and
  Parr}}]{zhao94}
\bibinfo{author}{\bibfnamefont{Q.}~\bibnamefont{Zhao}},
  \bibinfo{author}{\bibfnamefont{R.~C.} \bibnamefont{Morrison}},
  \bibnamefont{and} \bibinfo{author}{\bibfnamefont{R.~G.} \bibnamefont{Parr}},
  \bibinfo{journal}{Phys. Rev. A} \textbf{\bibinfo{volume}{50}},
  \bibinfo{pages}{238} (\bibinfo{year}{1995}).

\bibitem[{\citenamefont{Wu and Yang}(2003)}]{wuyang03}
\bibinfo{author}{\bibfnamefont{Q.}~\bibnamefont{Wu}} \bibnamefont{and}
  \bibinfo{author}{\bibfnamefont{W.}~\bibnamefont{Yang}}, \bibinfo{journal}{J.
  Chem. Phys.} \textbf{\bibinfo{volume}{118}}, \bibinfo{pages}{2498}
  (\bibinfo{year}{2003}).

\bibitem[{\citenamefont{de~Silva and Wesolowski}(2012)}]{weso12}
\bibinfo{author}{\bibfnamefont{P.}~\bibnamefont{de~Silva}} \bibnamefont{and}
  \bibinfo{author}{\bibfnamefont{T.~A.} \bibnamefont{Wesolowski}},
  \bibinfo{journal}{Phys. Rev. A} \textbf{\bibinfo{volume}{85}},
  \bibinfo{pages}{032518} (\bibinfo{year}{2012}).

\bibitem[{\citenamefont{Roncero et~al.}(2008)\citenamefont{Roncero,
  de~Lara-Castells, Villarreal, Flores, Ortega, Paniagua, and
  Aguado}}]{roncero08}
\bibinfo{author}{\bibfnamefont{O.}~\bibnamefont{Roncero}},
  \bibinfo{author}{\bibfnamefont{M.~P.} \bibnamefont{de~Lara-Castells}},
  \bibinfo{author}{\bibfnamefont{P.}~\bibnamefont{Villarreal}},
  \bibinfo{author}{\bibfnamefont{F.}~\bibnamefont{Flores}},
  \bibinfo{author}{\bibfnamefont{J.}~\bibnamefont{Ortega}},
  \bibinfo{author}{\bibfnamefont{M.}~\bibnamefont{Paniagua}}, \bibnamefont{and}
  \bibinfo{author}{\bibfnamefont{A.}~\bibnamefont{Aguado}},
  \bibinfo{journal}{J. Chem. Phys.} \textbf{\bibinfo{volume}{129}},
  \bibinfo{eid}{184104} (\bibinfo{year}{2008}).

\bibitem[{\citenamefont{Roncero et~al.}(2009)\citenamefont{Roncero, Zanchet,
  Villarreal, and Aguado}}]{roncero09}
\bibinfo{author}{\bibfnamefont{O.}~\bibnamefont{Roncero}},
  \bibinfo{author}{\bibfnamefont{A.}~\bibnamefont{Zanchet}},
  \bibinfo{author}{\bibfnamefont{P.}~\bibnamefont{Villarreal}},
  \bibnamefont{and} \bibinfo{author}{\bibfnamefont{A.}~\bibnamefont{Aguado}},
  \bibinfo{journal}{J. Chem. Phys.} \textbf{\bibinfo{volume}{131}},
  \bibinfo{eid}{234110} (\bibinfo{year}{2009}).

\bibitem[{\citenamefont{Goodpaster et~al.}(2010)\citenamefont{Goodpaster,
  Ananth, Manby, and {Miller III}}}]{goodpaster10}
\bibinfo{author}{\bibfnamefont{J.~D.} \bibnamefont{Goodpaster}},
  \bibinfo{author}{\bibfnamefont{N.}~\bibnamefont{Ananth}},
  \bibinfo{author}{\bibfnamefont{F.~R.} \bibnamefont{Manby}}, \bibnamefont{and}
  \bibinfo{author}{\bibfnamefont{T.~F.} \bibnamefont{{Miller III}}},
  \bibinfo{journal}{J. Chem. Phys.} \textbf{\bibinfo{volume}{133}},
  \bibinfo{eid}{084103} (\bibinfo{year}{2010}).

\bibitem[{\citenamefont{Goodpaster et~al.}(2011)\citenamefont{Goodpaster,
  Barnes, and {Miller III}}}]{goodpaster11}
\bibinfo{author}{\bibfnamefont{J.~D.} \bibnamefont{Goodpaster}},
  \bibinfo{author}{\bibfnamefont{T.~A.} \bibnamefont{Barnes}},
  \bibnamefont{and} \bibinfo{author}{\bibfnamefont{T.~F.} \bibnamefont{{Miller
  III}}}, \bibinfo{journal}{J. Chem. Phys.} \textbf{\bibinfo{volume}{134}},
  \bibinfo{eid}{164108} (\bibinfo{year}{2011}).

\bibitem[{\citenamefont{K\"ummel and Kronik}(2008)}]{kummel2008}
\bibinfo{author}{\bibfnamefont{S.}~\bibnamefont{K\"ummel}} \bibnamefont{and}
  \bibinfo{author}{\bibfnamefont{L.}~\bibnamefont{Kronik}},
  \bibinfo{journal}{Rev. Mod. Phys.} \textbf{\bibinfo{volume}{80}},
  \bibinfo{pages}{3} (\bibinfo{year}{2008}).

\bibitem[{\citenamefont{Jacob}(2011)}]{jacob11}
\bibinfo{author}{\bibfnamefont{C.~R.} \bibnamefont{Jacob}},
  \bibinfo{journal}{J. Chem. Phys.} \textbf{\bibinfo{volume}{135}},
  \bibinfo{eid}{244102} (\bibinfo{year}{2011}).

\bibitem[{\citenamefont{He{\ss}elmann et~al.}(2007)\citenamefont{He{\ss}elmann,
  G\"{o}tz, {Della Sala}, Manby, and G\"orling}}]{hesselman}
\bibinfo{author}{\bibfnamefont{A.}~\bibnamefont{He{\ss}elmann}},
  \bibinfo{author}{\bibfnamefont{A.~W.} \bibnamefont{G\"{o}tz}},
  \bibinfo{author}{\bibfnamefont{F.}~\bibnamefont{{Della Sala}}},
  \bibinfo{author}{\bibfnamefont{F.}~\bibnamefont{Manby}}, \bibnamefont{and}
  \bibinfo{author}{\bibfnamefont{A.}~\bibnamefont{G\"orling}},
  \bibinfo{journal}{J. Chem. Phys.} \textbf{\bibinfo{volume}{127}},
  \bibinfo{pages}{054102} (\bibinfo{year}{2007}).

\bibitem[{\citenamefont{Staroverov et~al.}(2006)\citenamefont{Staroverov,
  Scuseria, and Davidson}}]{staroverov}
\bibinfo{author}{\bibfnamefont{V.~N.} \bibnamefont{Staroverov}},
  \bibinfo{author}{\bibfnamefont{G.~E.} \bibnamefont{Scuseria}},
  \bibnamefont{and} \bibinfo{author}{\bibfnamefont{E.~R.}
  \bibnamefont{Davidson}}, \bibinfo{journal}{J. Chem. Phys.}
  \textbf{\bibinfo{volume}{124}}, \bibinfo{pages}{141103}
  (\bibinfo{year}{2006}).

\bibitem[{\citenamefont{Heaton-Burgess
  et~al.}(2007)\citenamefont{Heaton-Burgess, Bulat, and Yang}}]{heatonburgess}
\bibinfo{author}{\bibfnamefont{T.}~\bibnamefont{Heaton-Burgess}},
  \bibinfo{author}{\bibfnamefont{F.~A.} \bibnamefont{Bulat}}, \bibnamefont{and}
  \bibinfo{author}{\bibfnamefont{W.}~\bibnamefont{Yang}},
  \bibinfo{journal}{Phys. Rev. Lett.} \textbf{\bibinfo{volume}{98}},
  \bibinfo{pages}{256401} (\bibinfo{year}{2007}).

\bibitem[{\citenamefont{Yang et~al.}(1986)\citenamefont{Yang, Parr, and
  Lee}}]{yang86}
\bibinfo{author}{\bibfnamefont{W.}~\bibnamefont{Yang}},
  \bibinfo{author}{\bibfnamefont{R.~G.} \bibnamefont{Parr}}, \bibnamefont{and}
  \bibinfo{author}{\bibfnamefont{C.}~\bibnamefont{Lee}},
  \bibinfo{journal}{Phys. Rev. A} \textbf{\bibinfo{volume}{34}},
  \bibinfo{pages}{4586} (\bibinfo{year}{1986}).

\bibitem[{\citenamefont{García-Aldea and Alvarellos}(2007)}]{alva07}
\bibinfo{author}{\bibfnamefont{D.}~\bibnamefont{García-Aldea}}
  \bibnamefont{and} \bibinfo{author}{\bibfnamefont{J.~E.}
  \bibnamefont{Alvarellos}}, \bibinfo{journal}{J. Chem. Phys.}
  \textbf{\bibinfo{volume}{127}}, \bibinfo{eid}{144109} (\bibinfo{year}{2007}).

\bibitem[{\citenamefont{Ayers et~al.}(2002)\citenamefont{Ayers, Parr, and
  Nagy}}]{ayers02}
\bibinfo{author}{\bibfnamefont{P.~W.} \bibnamefont{Ayers}},
  \bibinfo{author}{\bibfnamefont{R.~G.} \bibnamefont{Parr}}, \bibnamefont{and}
  \bibinfo{author}{\bibfnamefont{A.}~\bibnamefont{Nagy}},
  \bibinfo{journal}{Int. Jou. Quant. Chem.} \textbf{\bibinfo{volume}{90}},
  \bibinfo{pages}{309} (\bibinfo{year}{2002}).

\bibitem[{\citenamefont{Thomas}(1926)}]{thomas26}
\bibinfo{author}{\bibfnamefont{L.~H.} \bibnamefont{Thomas}},
  \bibinfo{journal}{Proc. Cambridge Phil. Soc.} \textbf{\bibinfo{volume}{23}},
  \bibinfo{pages}{542} (\bibinfo{year}{1926}).

\bibitem[{\citenamefont{Fermi}(1928)}]{fermi28}
\bibinfo{author}{\bibfnamefont{E.}~\bibnamefont{Fermi}},
  \bibinfo{journal}{Rend. Accad. Naz. Lincei} \textbf{\bibinfo{volume}{48}},
  \bibinfo{pages}{73} (\bibinfo{year}{1928}).

\bibitem[{\citenamefont{Fermi}(1927)}]{fermi27}
\bibinfo{author}{\bibfnamefont{E.}~\bibnamefont{Fermi}}, \bibinfo{journal}{Z.
  Phys.} \textbf{\bibinfo{volume}{6}}, \bibinfo{pages}{602}
  (\bibinfo{year}{1927}).

\bibitem[{\citenamefont{{von Weizs\"{a}cker}}(1935)}]{vw}
\bibinfo{author}{\bibfnamefont{C.~F.} \bibnamefont{{von Weizs\"{a}cker}}},
  \bibinfo{journal}{Z. Phys. A} \textbf{\bibinfo{volume}{96}},
  \bibinfo{pages}{431} (\bibinfo{year}{1935}).

\bibitem[{\citenamefont{{Della Sala} et~al.}(2015)\citenamefont{{Della Sala},
  Fabiano, and Constantin}}]{alpha}
\bibinfo{author}{\bibfnamefont{F.}~\bibnamefont{{Della Sala}}},
  \bibinfo{author}{\bibfnamefont{E.}~\bibnamefont{Fabiano}}, \bibnamefont{and}
  \bibinfo{author}{\bibfnamefont{L.~A.} \bibnamefont{Constantin}},
  \bibinfo{journal}{Phys. Rev. B} \textbf{\bibinfo{volume}{91}},
  \bibinfo{pages}{035126} (\bibinfo{year}{2015}).

\bibitem[{\citenamefont{Brack et~al.}(1976)\citenamefont{Brack, Jennings, and
  Chu}}]{brack1976}
\bibinfo{author}{\bibfnamefont{M.}~\bibnamefont{Brack}},
  \bibinfo{author}{\bibfnamefont{B.}~\bibnamefont{Jennings}}, \bibnamefont{and}
  \bibinfo{author}{\bibfnamefont{Y.}~\bibnamefont{Chu}},
  \bibinfo{journal}{Phys. Lett. B} \textbf{\bibinfo{volume}{65}},
  \bibinfo{pages}{1 } (\bibinfo{year}{1976}).

\bibitem[{\citenamefont{Kirzhnits}(1957)}]{kirzhnits57}
\bibinfo{author}{\bibfnamefont{D.~A.} \bibnamefont{Kirzhnits}},
  \bibinfo{journal}{Sov. Phys. JETP} \textbf{\bibinfo{volume}{5}},
  \bibinfo{pages}{64} (\bibinfo{year}{1957}).

\bibitem[{\citenamefont{Lee et~al.}(2009)\citenamefont{Lee, Constantin, Perdew,
  and Burke}}]{mge2}
\bibinfo{author}{\bibfnamefont{D.}~\bibnamefont{Lee}},
  \bibinfo{author}{\bibfnamefont{L.~A.} \bibnamefont{Constantin}},
  \bibinfo{author}{\bibfnamefont{J.~P.} \bibnamefont{Perdew}},
  \bibnamefont{and} \bibinfo{author}{\bibfnamefont{K.}~\bibnamefont{Burke}},
  \bibinfo{journal}{J. Chem. Phys.} \textbf{\bibinfo{volume}{130}},
  \bibinfo{eid}{034107} (\bibinfo{year}{2009}).

\bibitem[{\citenamefont{Perdew et~al.}(1996{\natexlab{a}})\citenamefont{Perdew,
  Burke, and Ernzerhof}}]{pbe}
\bibinfo{author}{\bibfnamefont{J.~P.} \bibnamefont{Perdew}},
  \bibinfo{author}{\bibfnamefont{K.}~\bibnamefont{Burke}}, \bibnamefont{and}
  \bibinfo{author}{\bibfnamefont{M.}~\bibnamefont{Ernzerhof}},
  \bibinfo{journal}{Phys. Rev. Lett.} \textbf{\bibinfo{volume}{77}},
  \bibinfo{pages}{3865} (\bibinfo{year}{1996}{\natexlab{a}}).

\bibitem[{\citenamefont{Adamo and Barone}(1999)}]{pbe0_1}
\bibinfo{author}{\bibfnamefont{C.}~\bibnamefont{Adamo}} \bibnamefont{and}
  \bibinfo{author}{\bibfnamefont{V.}~\bibnamefont{Barone}},
  \bibinfo{journal}{J. Chem. Phys.} \textbf{\bibinfo{volume}{110}},
  \bibinfo{pages}{6158} (\bibinfo{year}{1999}).

\bibitem[{\citenamefont{Perdew et~al.}(1996{\natexlab{b}})\citenamefont{Perdew,
  Ernzerhof, and Burke}}]{pbe0_2}
\bibinfo{author}{\bibfnamefont{J.~P.} \bibnamefont{Perdew}},
  \bibinfo{author}{\bibfnamefont{M.}~\bibnamefont{Ernzerhof}},
  \bibnamefont{and} \bibinfo{author}{\bibfnamefont{K.}~\bibnamefont{Burke}},
  \bibinfo{journal}{J. Chem. Phys.} \textbf{\bibinfo{volume}{105}},
  \bibinfo{pages}{9982} (\bibinfo{year}{1996}{\natexlab{b}}).

\bibitem[{\citenamefont{Weigend and Ahlrichs}(2005)}]{def2tzvpp}
\bibinfo{author}{\bibfnamefont{F.}~\bibnamefont{Weigend}} \bibnamefont{and}
  \bibinfo{author}{\bibfnamefont{R.}~\bibnamefont{Ahlrichs}},
  \bibinfo{journal}{Phys. Chem. Chem. Phys.} \textbf{\bibinfo{volume}{7}},
  \bibinfo{pages}{3297} (\bibinfo{year}{2005}).

\bibitem[{\citenamefont{Rappoport and Furche}(2010)}]{furchepol}
\bibinfo{author}{\bibfnamefont{D.}~\bibnamefont{Rappoport}} \bibnamefont{and}
  \bibinfo{author}{\bibfnamefont{F.}~\bibnamefont{Furche}},
  \bibinfo{journal}{J. Chem. Phys.} \textbf{\bibinfo{volume}{133}},
  \bibinfo{pages}{134105} (\bibinfo{year}{2010}).

\bibitem[{tur()}]{turbomole}
\bibinfo{note}{{TURBOMOLE V6.2, 2009}, a development of {University of
  Karlsruhe} and {Forschungszentrum Karlsruhe GmbH}, 1989-2007, {TURBOMOLE
  GmbH}, since 2007; available from {\tt http://www.turbomole.com}.}

\bibitem[{\citenamefont{Zhao and Truhlar}(2005{\natexlab{a}})}]{truhlar05a}
\bibinfo{author}{\bibfnamefont{Y.}~\bibnamefont{Zhao}} \bibnamefont{and}
  \bibinfo{author}{\bibfnamefont{D.~G.} \bibnamefont{Truhlar}},
  \bibinfo{journal}{J. Phys. Chem. A} \textbf{\bibinfo{volume}{109}},
  \bibinfo{pages}{5656} (\bibinfo{year}{2005}{\natexlab{a}}).

\bibitem[{\citenamefont{Zhao and Truhlar}(2005{\natexlab{b}})}]{truhlar05nb}
\bibinfo{author}{\bibfnamefont{Y.}~\bibnamefont{Zhao}} \bibnamefont{and}
  \bibinfo{author}{\bibfnamefont{D.~G.} \bibnamefont{Truhlar}},
  \bibinfo{journal}{J. Chem. Theory Comput.} \textbf{\bibinfo{volume}{1}},
  \bibinfo{pages}{415} (\bibinfo{year}{2005}{\natexlab{b}}).

\end{thebibliography}
\end{document}